\documentclass[12pt,reqno]{amsart}

\usepackage{amsmath,amsfonts,amsthm,amssymb}

\usepackage[letterpaper,height=20cm,top=24mm,bottom=24mm,left=26mm,right=26mm]{geometry}

\usepackage[numbers,sort]{natbib}
\usepackage[ps2pdf]{hyperref}

\usepackage[footnotesize,hang,
]{caption}\setcaptionwidth{12cm}

\usepackage[dvips]{graphicx}
\usepackage{manfnt}

\usepackage{times}

\newcommand{\I}{\mathrm{i}}
\newcommand{\E}{\mathrm{e}}
\usepackage{titlesec}
\titleformat{\section}
{\bfseries\scshape\centering}
{\thesection.}{.5em}{}
\titleformat{\subsection}
{\rmfamily\bfseries}
{\thesubsection}{.5em}{}
\titleformat{\subsubsection}
{\rmfamily\bfseries}
{\manstar}{.5em}{}
\numberwithin{equation}{section}
\DeclareMathOperator{\Tr}{Tr}

\DeclareMathOperator{\diag}{diag}
\DeclareMathOperator{\Tet}{Tet}
\DeclareMathDelimiter{\Norm}{\mathord}{largesymbols}{"3E}{largesymbols}{"3E}
\begin{document}

%\special{header=psm.pro}

\renewcommand{\thefootnote}{\fnsymbol{footnote}}
%%%
\baselineskip 16pt
\parskip 8pt
\sloppy

% \noindent
% \UT
% \vskip -2cm
%\hfill  \today

%\vspace{27pt}

%%%%%%%%%%%%%%%%%% TITLE %%%%%%%%%%%%%%

\title[Skein Theory \&  Topological Quantum Registers]{
  Skein Theory and  Topological Quantum Registers:
  \\
  Braiding Matrices
  and 
  Topological Entanglement Entropy
  of Non-Abelian Quantum  Hall States
}

%%%%%%%%%%%%%%%%%%%%%%% AUTHOR(S) %%%%%%%%%%%%%%%%%%%

\author[K. Hikami]{Kazuhiro \textsc{Hikami}}

%%%%%%%%%%%%%%%%%%%%%%% ADDRESS %%%%%%%%%%%%%%%%%%%%

  \address{Department of Physics, Graduate School of Science,
%    \\
    University of Tokyo,
%    \\
    Hongo 7--3--1, Bunkyo, Tokyo 113--0033,   Japan.
    }
%     \\

%    \urladdr{http://gogh.phys.s.u-tokyo.ac.jp/{\textasciitilde}hikami/}
    \urladdr{\url{http://gogh.phys.s.u-tokyo.ac.jp/~hikami/}}

    \email{\texttt{hikami@phys.s.u-tokyo.ac.jp}}

%%%%%%%%%%%%%%%%%%%%%% DATE %%%%%%%%%%%%%%%%%%%%%%%%%%%
%(Received: \hspace{40mm})

%\vspace{18pt}
%\date{\today}
\date{September 14, 2007}

%%%%%%%%%%%%%%%%%%%%%% ABSTRACT %%%%%%%%%%%%%%%%%%%%%%
\begin{abstract}
We study topological properties of quasi-particle states 
in the  non-Abelian quantum Hall states.
We apply a skein-theoretic method to the Read--Rezayi state
whose effective theory is the $SU(2)_K$ Chern--Simons theory.
As a generalization of the Pfaffian ($K=2$) and the Fibonacci
($K=3$) anyon states,
we
compute the braiding matrices of quasi-particle states with arbitrary
spins.
Furthermore
we propose a method to compute the entanglement entropy
skein-theoretically.
We find that the entanglement entropy has a nontrivial contribution
called the topological entanglement entropy
which depends on the quantum 
dimension of  non-Abelian quasi-particle intertwining two
subsystems.
\end{abstract}

%%%%%%%%%%%%%%%%%%%%%%% Key Words %%%%%%%%%%%%%%%%%%%%%%%%%%

%\noindent
%\textsf{Key Words:}

%%\textsl{PACS:}
% \subjclass[2000]{
%   14J17,
%   32S25,
%   57M27
% }

%%%%%%%%%%%%%%%%%%%%%%%%%%%%%%%%%%%%%%%%%%%%%%%%%%%%%%%%%
%\newpage

%%\renewcommand{\thefootnote}{\arabic{footnote}}
\maketitle
%%%%%%%%%%%%%%%%%%%%%%%%%%%%%%%%%%%%

\section{Introduction}

Since the discovery of the fractional Hall effect and the Laughlin
wave function, it has been extensively
studied the fractional statistics, \emph{i.e.},
anyon (see \emph{e.g.} Ref.~\citenum{FWilc90Book}).
The effective theory of the Laughlin wave function for
$\nu=\frac{1}{2  \, m+1}$ is 
the $U(1)$ Chern--Simons (CS) theory~\cite{Read89,ZhanHansKive89a},
and those quasi-particles as excited
state
are understood as
the Abelian anyon.
Contrary to these fractional quantum  Hall states, the $\nu=5/2$ Hall state
experimentally observed in Ref.~\citenum{WESTGE87a}
remains mysterious
(see \emph{e.g.} Ref.~\citenum{NRead01a}).
One of theoretical  candidates for this state is the Moore--Read Pfaffian
state~\cite{MoRe91}, which
is related to the BCS ground state wave function of the
$p+ \I \, p$
superconductor such as Sr$_2$RuO$_4$~\cite{NReaGree99a,GreiXGWeWilc92a}.
The effective theory of the Pfaffian state was identified with the
$SU(2)_2$ CS theory~\cite{FradNayaTsveWilc98a},
and the non-Abelian property and the dimension of the Hilbert space of
the many-quasi-particle state have been studied from this
viewpoint~\cite{NayaWilc96a}.

As a generalization of the Pfaffian state as the non-Abelian quantum
Hall state,
proposed is the Read--Rezayi state~\cite{NReaERez99a}.
This state with $K=3$ is expected to describe the   $\nu=12/5$ quantum
Hall states reported in
Ref.~\citenum{XPVASSTPBW04}, and also pointed out is
a relationship  with the rotating Bose--Einstein
condensates~\cite{CoopWilkGunn01a}.
Originally the Read--Rezayi state was constructed 
by use of  the  $\mathbb{Z}_K$ parafermion theory~\cite{ZaFa85} which
is closely related
to the
$SU(2)_K$ CS theory.
%, and the case of $K=2$ reduces to the
%Pfaffian state.
% In fact this
% realize the fact in Ref.~\cite{NayaWilc96a} that $2n$-quasihole states
% of the Pfaffian state
% span $2^{n-1}$-dinmensional Hilbert space.
Algebraic structure of the Read--Rezayi state was analyzed by use of
the quantum group $U_q(SU(2))$, and the non-Abelian property as the braiding
matrices of quasi-particle states was studied~\cite{SlingFBais01a}.

The CS field theory plays an important role in the
quantum topological invariant of links and 3-manifolds.
Witten constructed the quantum invariant of  3-manifolds from the
Chern--Simons partition function with gauge group $G$ based on
conformal block~\cite{MoorSeib89a}, and he further
identified 
quantum polynomial invariants such as the Jones
polynomial~\cite{Jones85} and the HOMFLY polynomial~\cite{FYHLMO85}
as the expectation value of the Wilson loop of the CS
theory~\cite{EWitt89a}.
% shown in Ref.\citenum{EWitt89a} is that the CS path integral 
% The $SU(2)_K$ Chern--Simons (CS) theory gives the quantum invariant of
% links and 3-manifolds~
% The expectation value of the Wilson line coincides with the colored
% Jones polynomial.
His construction in terms of path integral
is not mathematically rigorous, and the invariant of 3-manifold was
later constructed combinatorially
by Reshetikhin and  Turaev~\cite{ResheTurae91a}, and
it  is called the Witten--Reshetikhin--Turaev (WRT)
invariant.

Recently from the viewpoint of the quantum computation
(see \emph{e.g.} Refs.~\citenum{NielsChuan00Book,KitaShenVyal02a}),
the
non-Abelian quantum Hall states have received renewed interests.
As a fault-tolerant quantum computing system, Kitaev proposed to
construct quantum registers topologically~\cite{AKitae03a,AKitae06a}.
Though his toric code uses a spin system on 2-dimensional honeycomb
lattice as
a non-Abelian quasi-particle system
(see also Ref.~\citenum{LeviXWen05a} where topological phase of
trivalent graph, sometimes called  string net or spin network, is
discussed),
the quantum Hall system is another
candidate for non-Abelian statistics
(see Refs.~\citenum{LGeorg06b,BondShteSlin07a,SarFreNaySimSte07a} for
recent reviews).
In such  topological quantum registers, unitary operations are constructed
from   braiding
of non-Abelian anyons.
It was shown~\cite{FreeLarsWang02a,FreeLarsWang02b} that the braiding
in the $SU(2)_K$ Chern--Simons theory  can efficiently approximate,
\emph{i.e.}, quantum compile,
any
quantum computing gate;
$n_{\geq 3}$-braid group for $K\geq 3$, $K\neq 4, 8$
and
$n_{\geq 5}$-braid group for $K=8$ 
are
capable of universal quantum computation.

One of our purposes in this article is to study the braiding property of
non-Abelian quasi-particles in the $SU(2)_K$ CS theory.
By use of explicit form of correlation function in
Ref.~\citenum{KniZam84},
Ardonne and Schoutens studied  braiding matrices~\cite{ArdonSchou07a}
for 4-quasi-particle states.
It becomes difficult to derive braiding matrices for
many-quasi-particle states by their method.
We rather use the Temperley--Lieb algebra or the skein theory,
which is elementary and combinatorial,
to derive the braiding matrices.
Although, a na\"{i}ve application of
the skein theory does not give a unitary representation.
We show explicitly how to construct the \emph{unitary} braiding matrices by
use of skein theory.
See also Ref.~\citenum{KauffLomon07a},
where proposed is  a different method to construct the
unitary operators from the skein theory by modifying  a weight of
trivalent vertex.
Our method is essentially equivalent to that of
Ref.~\citenum{SlingFBais01a}, once we know both
a construction of the
colored Jones polynomial based on representation theory of the quantum
group and a quantum group structure of the conformal field theory
(see \emph{e.g.} Refs.~\citenum{GomeRuizSeir96Book,KassRossTura97a}).
So most of methods in this article might be well known,
except our new
result is a computation of
the  entanglement entropy
(see Refs.~\citenum{AmiFazOstVed07a} for recent review)
in the framework of the skein theory.
%to get unitary braiding operators.
% This gives a physical interpretation
% of a setting of parameter
% $A=\I \, \E^{\frac{\pi \, \I}{2  (K+2)}}$~\eqref{A_set_i_times}
% in the
% skein theory.
In a topological phase such as non-Abelian states, it was
discussed~\cite{KitaePresk06a,LeviXWen06a}
that the
entanglement entropy has a topological part,
which is called the
topological entanglement entropy,
and that
it is expected 
to detect a topological order~\cite{WenNiu90a}.
Some numerical and analytical
studies have been done for the quantum dimer model in a
triangular lattice~\cite{FurukMisgu07a},
for the fractional quantum Hall states of the fermionic Laughlin
type~\cite{HaquZozuScho07a},
and for the quantum eight-vertex model~\cite{PapaRamaFrad07a}.
Here we show that the  entanglement entropy of the
state
$
  \left| \Psi \right\rangle
  =
  \sum_j p_j \,
  \left| \psi_{j} \right\rangle_A \otimes
  \left| \phi_{j} \right\rangle_B 
$
with 
$\sum_j \left| p_j \right|^2 = 1$
is given by
\begin{equation*}
  S_A
  =
  \sum_j 
  \left| p_j \right|^2 \log
  \left( \frac{d_j}{
      \left| p_j \right|^2 
    }
  \right)
\end{equation*}
which includes the topological entanglement entropy
\begin{equation*}
  S_A^{\text{topo}}
  =
  \sum_j 
  \left| p_j \right|^2 \log d_j
\end{equation*}
Here $d_j$ denotes the quantum dimension of quasi-particle which
connects subspaces $A$ and $B$.

A content of this article is constructed as follows.
In Section~\ref{sec:QHE}, we review the Read--Rezayi state.
We shall discuss a relationship with the $\mathbb{Z}_K$ parafermion
CFT, the $SU(2)_K$ Chern--Simons theory, and the $SU(2)$ topological
quantum invariants.
We further  explain basic facts about the Temperley--Lieb skein theory
following
Ref.~\citenum{KaufLins94Book}.
In Section~\ref{sec:entropy} we outline a method to compute the
entanglement entropy of quasi-particle states skein-theoretically.
We   define the topological entanglement entropy from the von Neumann
entropy.
In Section~\ref{sec:spin-half} we construct  a unitary representation of
braid operators on spin-$1/2$ state in the $SU(2)_K$ CS theory.
The $SU(2)_2$  theory 
corresponds to the Moore--Read
Pfaffian state.
We shall further compute the  entanglement entropy of these
states, and we show that it has a topological entropy which depends on
the quantum dimension of quasi-particle.
Section~\ref{sec:spin-1} is devoted to studies of spin-$1$
quasi-particle states in the $SU(2)_K$ theory.
A case of $K=3$ is known as the Fibonacci anyon model.
We construct the bases of quasi-particle states, and give explicit
forms of the braiding matrices.
We also calculate the topological entanglement entropy.
In Section~\ref{sec:different} we study a correlation function which
is composed from different species of quasi-particles as another
candidate of topological quantum registers.
We give an elementary treatment of skein theory up to
Section~\ref{sec:different}
as best as we can.
Later in Section~\ref{sec:6j}
we shall give a general setup for many-quasi-particle
states  in terms of the quantum $6j$ symbol.
The braiding matrices and the topological entanglement entropy are
discussed by use of the quantum $6j$ symbol.
The last section is devoted to concluding remarks and discussions.

% The $SU(2)_2$ CS theory does not have a universal quantum
% computation cability~\cite{FreeLarsWang02a,FreeLarsWang02b}.

% Discussed in~\cite{LeviXWen05a} 
% was an importance of a trivalent graph,
% or string-net,
% in 
% spin  model on 2-dimensional honeycomb lattice.

%%%
\section{Non-Abelian Quantum Hall State,
  CFT, and Quantum Invariants}
\label{sec:QHE}
\subsection{Parafermion Theory and the Read--Rezayi State}

The Read--Rezayi state is constructed by use of the $\mathbb{Z}_K$
parafermion theory~\cite{ZaFa85,GepQiu87a}
(see also \emph{e.g.} Ref.~\citenum{DiFrMathSene97}).
The $\mathbb{Z}_K$ parafermion theory is equivalent to $SU(2)_K/U(1)$
theory,
and we use the primary fields
$\Phi_m^\ell$ with
$ - \ell < m \leq \ell$
%$m\in \mathbb{Z}$
and
$\ell \in \{ 0,1, \dots, K\}$
whose conformal dimension
%$h_m^\ell$ of $\Phi_m^\ell$
is
\begin{equation*}
  h_m^\ell
  =
  \frac{
    \ell \, (\ell +2)
  }{
    4 \, (K+2)}
  -
  \frac{m^2}{4 \, K}
\end{equation*}
Here $m$ denotes the U(1) charge defined modulo $2 \, K$,
and
% we
% choose
% $-\ell < m \leq  \ell$
parameter $\ell$  parametrizes
a spin-$\ell/2$.
The primary fields $G_m^{(\ell)}$
with $\ell, m \in \mathbb{Z}/2$ of the $SU(2)_K$ theory is 
\begin{equation}
  \label{primary_WZW_parafermion}
  G_m^{(\ell)}
  =
  \Phi_{2m}^{2 \ell} \,
  \exp \left(
    \frac{\I \, m}{\sqrt{K}} \,
    \varphi
  \right)    
\end{equation}
where $\varphi$ is the free massless Bose field.
Further for the field $\Phi_m^\ell$
we have
$\ell=m \mod 2$ and
$\Phi_m^\ell = \Phi_{m+2 K}^\ell
=
\Phi_{m - K}^{K - \ell}$.
The fusion rule follows from the $SU(2)_K$ theory as
\begin{equation}
  \label{fusion_parafermion}
  \Phi_m^\ell \times
  \Phi_{m^\prime}^{\ell^\prime}
  =
  \sum_{
    \ell^{\prime \prime} =
    \left| \ell - \ell^\prime \right|
  }^{
    \min
    \left(
      \ell+\ell^\prime,
      2 K - \ell - \ell^\prime
    \right)
  }
  \Phi_{m+m^\prime}^{\ell^{\prime \prime}}
\end{equation}

In terms of the primary fields $\Phi_m^\ell$,
the vacuum sector $I$ is
$\Phi_0^0 = \Phi_K^K= I$, and
the
parafermions $\psi_\ell$,
the spin fields $\sigma_\ell$,
the dual spin fields $\mu_\ell$
% with $i=1,\dots, K$
and neutral fields $\varepsilon_\ell$
are respectively defined by
\begin{equation}
  \begin{gathered}
    \psi_\ell
    =
    \Phi_{2 \ell}^0
    =
    \Phi_{2 \ell -K}^K
    \\[2mm]
    \sigma_\ell
     =
    \Phi_\ell^\ell
    \\[2mm]
    \mu_\ell = \Phi_{-\ell}^\ell
    \\[2mm]
    \varepsilon_\ell
    =
    \Phi_0^{2 \ell}
  \end{gathered}
\end{equation}
and we mean
$\psi_\ell^\dagger = \psi_{K-\ell}$, and
$\sigma_\ell^\dagger = \sigma_{K-\ell}$.
Note that
$h_{\psi_\ell}=
\frac{
  \ell \, \left( K-\ell \right)}{
  K}
$,
$h_{\sigma_\ell}
= h_{\mu_\ell}
=\frac{\ell \, (K- \ell)}{2 \, K \, (K+2)}$,
and
$h_{\varepsilon_\ell}
=
\frac{\ell \, (\ell+1)}{K+2}$.

It is noted that the field $\psi_1$ for $K=2$ becomes the Majorana
fermion $\psi$ satisfying
$\left\langle \psi(z) \, \psi(w) \right\rangle = (z-w)^{-1}$,
and the Pfaffian state is identified with  the correlation function
$\left\langle
  \psi (z_1) \cdots \psi(z_N)
\right\rangle$~\cite{MoRe91}.
Correspondingly
the wave function of  the $n$-quasi-particle states can be constructed from
the correlation function
$\left\langle
  \sigma_1(w_1) \cdots \sigma_1(w_n) \,
  \psi(z_1) \cdots \psi(z_N)
\right\rangle$.
This function  was  studied in detail~\cite{NayaWilc96a}, and 
the dimension of the wave function for
$2 \, n$-quasi-particle states are
identified to be 
$2^{n-1}$.

The Read--Rezayi state~\cite{NReaERez99a} is  a generalization of the
Pfaffian state, and
the wave function  is defined as correlation
function of the primary fields in
the  $\mathbb{Z}_K$ parafermion theory;
\begin{multline}
  \label{Read_Rezayi_state}
  \Psi(w_1,\dots, w_n ; z_1, \dots , z_N)
  =
  \left\langle
    \sigma_1(w_1) \cdots \sigma_1(w_n) \,
    \psi_1(z_1) \cdots \psi_1(z_N)
  \right\rangle \\
  \times
  \prod_{i<j} \left( z_i - z_j \right)^{M+\frac{2}{K}}
  \,
  \prod_{i=1}^N \prod_{j=1}^n
  \left(z_i - w_j \right)^{\frac{1}{K}} \,
  \prod_{i<j} 
  \left(w_i - w_j \right)^{\frac{1}{K (K M+2)}} \,
%   \\
%   \times
%   F_g(w_1,\dots,w_n; z_1, \dots, z_N)
\end{multline}
Here $M$ is an odd or even integer depending on  whether the
particles are fermions or bosons, and the filling factor becomes
\begin{equation*}
  \nu = \frac{K}{K \, M+2}
\end{equation*}
The fractional power in the second line of~\eqref{Read_Rezayi_state}
was added to cancel the singular power which comes from the operator
product expansion of $\Phi_m^\ell$ so that the wave function is
non-singular.
We should note
that the Read--Rezayi state~\eqref{Read_Rezayi_state} without
quasi-particles
($n=0$)  is an exact ground state of the Hamiltonian with ($K+1$)-body
interactions;
\begin{equation}
  \mathcal{H}
  =
  V \,
  \sum_{i_1 < i_2 \cdots < i_{K+1}}
  \delta^2(z_{i_1} - z_{i_2}) \,
  \delta^2(z_{i_2} - z_{i_3})
  \cdots
  \delta^2(z_{i_K} - z_{i_{K+1}})
\end{equation}

So to study  the action of braiding  of quasi-particles on the
wave function~\eqref{Read_Rezayi_state},
we need the braiding actions on the correlation functions of parafermion
fields.
By use of correspondence~\eqref{primary_WZW_parafermion} with the
$SU(2)_K$ theory, 
we see that the braiding matrices coincide with that of the $SU(2)_K$
theory.

%%%
\subsection{Quantum Invariant from the Chern--Simons Theory}
In Ref.~\citenum{EWitt89a}, the physical interpretation of the quantum
knot invariant such as the Jones polynomial~\cite{Jones85}
and the HOMFLY polynomial~\cite{FYHLMO85} was given.
The Chern--Simons action with gauge group $G$ on 3-manifold
$\mathcal{M}$ is defined by
\begin{equation}
  \label{Chern-Simons_action}
  S_{\text{CS}}=\frac{K}{4 \, \pi}
  \int\limits_{\mathcal{M}}
  \Tr
  \left(
    A \wedge \mathrm{d} A
    + \frac{2}{3} \,
    A \wedge A \wedge A
  \right)
\end{equation}
where $K$ is a coupling constant $K\in \mathbb{Z}$, and $A$ is the
gauge connection on $\mathcal{M}$.
Witten introduced the topological invariant of $\mathcal{M}$ as the
partition function
\begin{equation}
  Z(\mathcal{M})
  =
  \int \mathcal{D} A \,
  \E^{\I \, S_{\text{CS}}}
\end{equation}
This invariant
was later mathematically rigorously defined by
Reshetikhin and Turaev~\cite{ResheTurae91a}, and it is called the
Witten--Reshetikhin--Turaev (WRT) invariant.
Simpler construction was given later in Ref.~\citenum{Licko93a} based
on the Conway--Kauffman skein relation (see also
Refs.~\citenum{KaufLins94Book,Licko97Book}).

Not only the invariant of the 3-manifold $\mathcal{M}$, the Chern--Simons
action~\eqref{Chern-Simons_action}
gives the quantum invariant for link
$\mathcal{L}$ in $\mathcal{M}$.
We define the Wilson loop expectation value by
\begin{equation}
  W_{R_1, \dots, R_L}(\mathcal{L})
  =
  \frac{1}{Z(\mathcal{M})}
  \int \mathcal{D} A \,
  W_{R_1}^{\mathcal{K}_1}(A) \cdots
  W_{R_L}^{\mathcal{K}_L}(A)
  \,
  \E^{\I \, S_{\text{CS}}}
\end{equation}
where we assume the link $\mathcal{L}$ has components $\mathcal{K}_1,
\dots
\mathcal{K}_L$, and $W_R^{\mathcal{K}}$ denotes the Wilson loop
operator along a component $\mathcal{K}$ with a representation $R$ of
gauge group $G$;
\begin{equation}
  W_R^{\mathcal{K}}(A)
  = \Tr_R
  \mathcal{P} \,
  \exp \oint\limits_\mathcal{K} A
\end{equation}
where $\mathcal{P}$ is a path ordering.
When the gauge group $G=SU(2)$, this correlation function becomes the
colored Jones polynomial at the root of unity.

%%%%%%%%%%%%%%%%%%%%%%%%%
\subsection{Combinatorial Construction of  $SU(2)$ Quantum Invariants}

Hereafter in this article  we only study a case of the gauge group
$G=SU(2)$.
In $SU(2)$ case, as a $S$-transformation of the conformal block, we
use~\cite{GeWi86}
\begin{equation}
  \label{S-matrix}
  S_{a b}
  =
  \sqrt{
    \frac{2}{K+2}} \,
  \sin
  \left(
    \frac{
      \left( a+1 \right) \,
      \left( b+1 \right) \,
      \pi
    }{K+2}
  \right)
\end{equation}
In actual computation of the quantum invariants such as
$Z(\mathcal{M})$ and $W_R(\mathcal{K})$, 
useful and the most elementary method, at least to the author,
is to  use the skein theory.
Here
we briefly review a skein-theoretical construction of the $SU(2)$
quantum invariant
(see, \emph{e.g.}, Ref.~\citenum{KaufLins94Book,Licko97Book}).

The Jones polynomial is characterized by the skein relation;
\begin{gather}
  \label{skein}
  \mbox{
    \raisebox{-.2cm}{
      \includegraphics[scale=0.7]{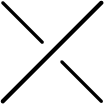}
    }
  }
  =
  A \,
  \mbox{
    \raisebox{-.2cm}{
      \includegraphics[scale=0.7]{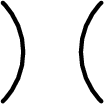}
    }
  }
  +
  A^{-1} \, 
  \mbox{
    \raisebox{-.2cm}{
      \includegraphics[scale=0.7]{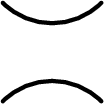}
    }
  }
  \\[2mm]
  \label{skein2}
  D \cup 
  \mbox{
    \raisebox{-.2cm}{
      \includegraphics[scale=0.6]{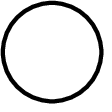}
    }
  }
  =
  d \, D
\end{gather}
Here $A$ is a parameter related to the quantum group
deformation parameter
$q=A^4$, and
\begin{equation}
  \label{define_d}
  d=-A^2 - A^{-2}    
\end{equation}
Projection of knots and links is denoted by $D$.
%Diagram $D$ is a projection of knot..

In constructing
the colored Jones polynomial of links and the  SU(2) WRT
invariant of
3-manifolds,
fundamental tool is the Jones--Wenzl idempotent.
Throughout this article, we employ a standard notation in diagrams $D$;
an integer $n$ beside an arc means that there exist $n$ copies of that
arc.
Like~\eqref{skein} an arc without integer denotes a single arc,
\emph{i.e.}, $n=1$ is often omitted. 
Then
the Jones--Wenzl idempotent is depicted as a blank square, and it may
also be
recursively defined as
\begin{equation}
  \mbox{
    \raisebox{-8mm}{
      \includegraphics[scale=0.88]{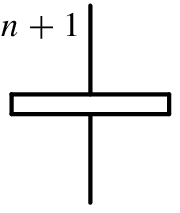}
    }
  }
  =
  \mbox{
    \raisebox{-8mm}{
      \includegraphics[scale=0.88]{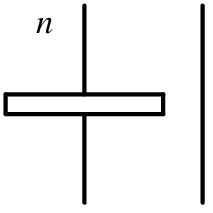}
    }
  }
  -
  \frac{\Delta_{n-1}}{\Delta_n} \,
  \mbox{
    \raisebox{-8mm}{
      \includegraphics[scale=0.88]{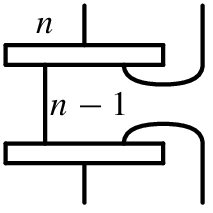}
    }
  }
  \label{Jones-Wenzl}
\end{equation}
Here $\Delta_n$ is the colored Jones polynomial for unknot;
\begin{equation}
  \begin{aligned}
    \Delta_n
    & =
    \mbox{
      \raisebox{-.6cm}{
        \includegraphics[scale=0.88]{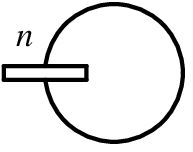}
      }
    }
    \\
    & =
    (-1)^n \frac{A^{2 n +2}-A^{-2n-2}}{A^2-A^{-2}}
  \end{aligned}
\end{equation}
We have
$  \Delta_0 = 1$,
$  \Delta_1 = d$,
and others are recursively defined by
\begin{gather}
  \Delta_{n+1} = d \, \Delta_n - \Delta_{n-1}
\end{gather}
Note that we have
\begin{gather*}
  \Delta_{x+y+z+1} \, \Delta_{z-1}
  =
  \Delta_{x+z} \, \Delta_{y+z} -
  \Delta_{y} \, \Delta_x
%   \left( \Delta_n \right)^2 -1
%   =
%   \Delta_{n+1} \, \Delta_{n-1}
\end{gather*}
One notices that the Jones--Wenzl idempotent satisfies followings;
\begin{gather}
  \mbox{
    \raisebox{-8mm}{
      \includegraphics[scale=0.88]{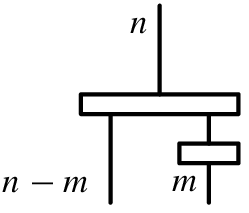}
    }
  }
  =
  \mbox{
    \raisebox{-8mm}{
      \includegraphics[scale=0.88]{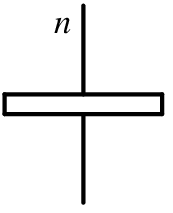}
    }
  }
  \\[2mm]
  \label{projection_zero}
  \mbox{
    \raisebox{-8mm}{
      \includegraphics[scale=0.88]{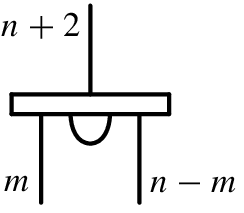}
    }
  }
  = 0
\end{gather}
where $n \geq m$.

In the $SU(2)_K$ theory, we have $\Delta_n \neq 0$ for
$1\leq n\leq K$, and $\Delta_{K+1}=0$;
namely the parameter $A$ satisfies
\begin{equation}
%  A=\exp \left( \frac{\pi \, \I}{2 \, (K+2)} \right)
  A^{4 (K+2)} = 1
\end{equation}
This condition means that
integer $n$ attached to each arc should be
$1 \leq n \leq K$.
Here we choose~\cite{FrNaShWaWa04a}
\begin{equation}
  \label{A_set_i_times}
  A= \I \, \exp \left(
    \frac{\pi \I}{2 \, (K+2)}
  \right)
\end{equation}
from physical requirement which will be explained below.
Under this condition we have
\begin{equation}
  \label{Delta_special}
  \Delta_n =
  \frac{\sin \left( \frac{n+1}{K+2} \, \pi \right)}{
    \sin \left( \frac{1}{K+2} \, \pi \right)}
\end{equation}
and
especially we have
% also~\eqref{define_d}
\begin{equation}
  d= \Delta_1 = 2 \cos \left( \frac{\pi}{K+2} \right)
\end{equation}

% The fusion relation is written in terms of the Jones--Wenzl
% idempotent.
We introduce a trivalent vertex by
\begin{equation}
  \label{trivalent}
  \mbox{
    \raisebox{-8mm}{
      \includegraphics[scale=0.88]{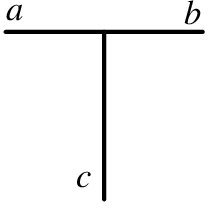}
    }
  }
  =
  \mbox{
    \raisebox{-8mm}{
      \includegraphics[scale=0.88]{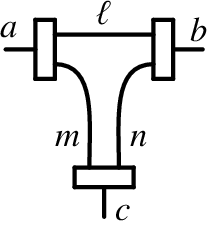}
    }
  }
\end{equation}
where
\begin{align*}
  \ell & = \frac{a+b-c}{2}
  &
  m & = \frac{c+a-b}{2}
  &
  n & = \frac{b+c-a}{2}
\end{align*}
Indices in trivalent vertex~\eqref{trivalent} should fulfill an  admissible
condition;
\begin{equation}
  \label{admissible}
  \begin{cases}
    a+b+c = 0 \mod 2
    \\[2mm]
    a+b\geq c,
    \qquad
    b+c\geq a,
    \qquad
    c+a\geq b
    \\[2mm]
    a+b+c \leq 2 \, K
  \end{cases}
\end{equation}

Skein relation~\eqref{skein} gives
\begin{equation}
  \begin{aligned}
    \mbox{
      \raisebox{-.7cm}{
        \includegraphics[scale=0.88]{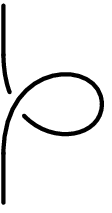}
      }
    }
    & =
    -A^3 \,
    \mbox{
      \raisebox{-.7cm}{
        \includegraphics[scale=0.88]{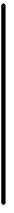}
      }
    }
    & \hspace{24mm}
    \mbox{
      \raisebox{-.7cm}{
        \includegraphics[scale=0.88]{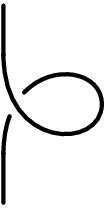}
      }
    }
    & =
    -A^{-3} \,
    \mbox{
      \raisebox{-.7cm}{
        \includegraphics[scale=0.88]{twist.0.eps}
      }
    }
  \end{aligned}
  \label{twist_simple}
\end{equation}
Generally we have a twist formula;
\begin{equation}
  \label{twist_formula}
  \mbox{
    \raisebox{-.7cm}{
      \includegraphics[scale=0.88]{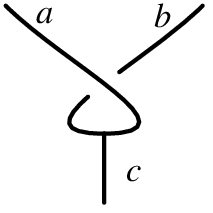}
    }
  }
  =
%   (-1)^{\frac{a+b-c}{2}} \,
%   A^{
%     a+b-c+\frac{a^2+b^2-c^2}{2}
%   } \,
  \lambda_c^{a  b} \,
  \mbox{
    \raisebox{-.7cm}{
      \includegraphics[scale=0.88]{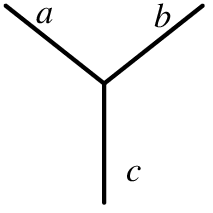}
    }
  }
\end{equation}
where
\begin{equation}
  \lambda^{a b}_c
  =
  (-1)^{\frac{a+b-c}{2}} \,
  A^{
    a+b-c+\frac{a^2+b^2-c^2}{2}
  } \,
\end{equation}

%%%%%%%%%
\subsection{Trivalent Diagram as Wilson Line of Quasi-Particles}

The diagram introduced above can be interpreted physically
as follows.
We regard a single arc as a Wilson line of quasi-particle with
spin-$1/2$.
The Jones--Wenzl idempotent which has index $n$ means
the projection to spin-$n/2$ space, and
the arc labelled by $n$ corresponds to the Wilson line of
quasi-particle with
spin-$n/2$;
\begin{equation*}
  \mbox{
    \raisebox{-.8cm}{
      \includegraphics[scale=0.88]{proj1.12.eps}
    }
  }
  :
  \text{spin-$n/2$ quasi-particles}
\end{equation*}
Then the trivalent vertex~\eqref{trivalent} denotes the usual
fusion rule;
quasi-particles, $\phi_a$ and $\phi_b$, with spin-$a/2$ and
spin-$b/2$ fuse to spin-$c/2$ quasi-particle $\phi_c$.
One sees that the admissible condition~\eqref{admissible}
of the $SU(2)_K$ CS theory   has a
correspondence with the fusion
channel~\eqref{fusion_parafermion}~\cite{SlingFBais01a}.
In general the fusion rule is written as
\begin{equation}
  \phi_a \times \phi_b =
  \sum_c
  N_{a b}^c \,
  \phi_c
\end{equation}
Verlinde formula~\cite{Verli88} shows that the
the fusion multiplicity can be written in terms of the $S$-matrix as
\begin{equation}
  N_{a b c}
  =
  \sum_d
  \frac{
    S_{a d} \, S_{b d} \, S_{c d}
  }{
    S_{0 d}
  }
\end{equation}
We recall that the 
$S$-matrix for $SU(2)$ is given in~\eqref{S-matrix}.
The quantum dimension $d_a$ of  quasi-particle with spin $a/2$ is the
largest eigenvalue of
the matrix $N_a$, and in the $SU(2)_K$ case we have
\begin{equation}
  \label{quantum_dimension}
  d_a 
  =
  \frac{S_{0 a}}{S_{0 0}}
  = \Delta_a
\end{equation}
% As we have seen before, the braiding matrices  of quasiparticles
% $\sigma_n$
% can
% be derived from the $SU(2)_K$ theory.
% The fusion rule~\eqref{fusion_parafermion}  can be read as
% \begin{equation*}
%   \sigma_a \times \sigma_b = \sum_{c=|a-b|}^{\min(a+b,2K-a-b)}
%   \sigma_c
% \end{equation*}
% and 
% identified with the admissible condition~\eqref{admissible} of the
% trivalent vertex.

The non-vanishing correlation function of quasi-particles should fuse
into the vacuum section.
For example, the correlation function of two quasi-particles with
spin-$1/2$ 
will vanish unless two fuse into the vacuum.
Diagrammatically we denote this 2-quasi-particle state by
\begin{equation}
  \left| \widetilde{\text{qp}} \right\rangle
  =
  \mbox{
    \raisebox{-.1cm}{
      \includegraphics[scale=0.8]{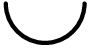}
    }
  }
\end{equation}
Two  endpoints of arc indicate existence of  quasi-particle at
spatially different  points, and  a connecting
arc denotes that they fuse to the
vacuum sector.
Naturally  a  dual state  is defined upside down as
\begin{equation}
  \left\langle \widetilde{\text{qp}} \right|
  =
  \mbox{
    \raisebox{-.1cm}{
      \includegraphics[scale=0.8]{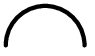}
    }
  }
\end{equation}
We can compute  norm of the state by connecting two endpoints of 
$\left\langle \widetilde{\text{qp}} \right|$ and
$\left| \widetilde{\text{qp}} \right\rangle$,
and~\eqref{skein2} gives
\begin{equation}
  \left\langle \widetilde{\text{qp}} |
    \widetilde{\text{qp}}
  \right\rangle
  =
  \mbox{
    \raisebox{-.3cm}{
      \includegraphics[scale=0.8]{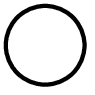}
    }
  }
  = d
\end{equation}
As we have set a parameter $A$ as~\eqref{A_set_i_times}, we have $d>0$
as a level
$K$   is
$K\geq 1$.
Then we have the normalized physical state,
which is schematically written as
\begin{equation}
  \label{normalized_qp}
  \left| \text{qp}
  \right\rangle
  =
  \frac{1}{\sqrt{d}} \,
  \mbox{
    \raisebox{-.1cm}{
      \includegraphics[scale=0.8]{qubit.81.eps}
    }
  }
\end{equation}

%%%%%%%%%%%%%%%%%%%%%%%%%
\section{Topological Entanglement Entropy}
\label{sec:entropy}

Entanglement is one of  distinguishing properties of quantum
mechanics, and it
receives much interests in recent studies of quantum information science.
One of   measures of entanglement is the entropy
(see, \emph{e.g.},
Refs.~\citenum{AmiFazOstVed07a} for recent review).
Namely when 
we assume that the state is a pure bipartite state
$\left| \Psi \right\rangle$, and that the
system is divided into two sub-systems $A$ and $B$,
the  Schmidt decomposition assures that we can write
\begin{equation}
  \label{pure_Psi}
  \left| \Psi \right\rangle
  =
  \sum_j p_j \,
  \left| \psi_{j} \right\rangle_A \otimes
  \left| \phi_{j} \right\rangle_B 
\end{equation}
where
$\left|\psi_{j} \right\rangle_A$
and
$\left|\phi_{j} \right\rangle_B$
are orthonormal states in subspace $A$ and $B$,
\begin{align}
  \label{orthogonal_base}
  {
    \vphantom{\left\langle \psi_j \right|}
  }_A \!
  \left\langle \psi_j \middle| \psi_k \right\rangle_A
  & =
  \delta_{j,k}
  &
  {
    \vphantom{\left\langle \phi_j \right|}
  }_B \!
  \left\langle \phi_j \middle| \phi_k \right\rangle_B
  & =
  \delta_{j,k}
\end{align}
and
$\sum_j
\left| p_j \right|^2 = 1$.
Then the entanglement entropy is defined by
\begin{equation}
  \label{original_entropy}
  S_A=
  - \sum_j
  \left| p_j \right|^2
  \log
  \left| p_j \right|^2
\end{equation}
We note that
\begin{equation}
  \label{entropy_A_B}
  S_A= S_B
\end{equation}

Alternatively the entanglement entropy is defined as the von Neumann
entropy
\begin{equation}
  \label{von_Neumann_entropy}
  S_A =
  - \Tr_A
  \left(
    \rho_A \, \log \rho_A
  \right)
\end{equation}
Here $\rho_A$ is the reduced density matrix
(Alice's density matrix)
\begin{equation}
  \label{reduced_density}
  \rho_A
  = \Tr_B \rho
  =
  \Tr_B
  \left(
    \left| \Psi \right\rangle \,
    \left\langle \Psi \right|
  \right)
\end{equation}
where the density matrix $\rho$ is
\begin{equation}
  \label{density_matrix}
  \rho= \left| \Psi \right\rangle \, \left\langle \Psi \right|
\end{equation}
% We defined the topological entanglement entropy as
% follows~\cite{KitaePresk06a,LeviXWen06a}.
In fact when we use the replica
trick~\cite{HolzLarsWilc94a,CalabCardy04b}
\begin{equation}
  \label{entropy_replica}
  S_A=
  - \lim_{n\to 1}
  \frac{\partial}{\partial n} \,
  \left(
    \Tr_A \rho_A^{~n}
  \right)
\end{equation}
we obtain~\eqref{original_entropy}.

In the non-Abelian  states, it is expected that there exists an
effect from a
topological order~\cite{KitaePresk06a,LeviXWen06a}.
Those papers discuss that such effect appears from a dependence on a
length of boundary between $A$ and $B$.
As our model is purely topological and pure gauge theory,
we rather  define the topological entanglement entropy by
\begin{equation}
  \label{topological_entanglement}
  S_A^{\text{topo}}
  =
%  S_A
  - \Tr_A
  \left(
    \rho_A \, \log \rho_A
  \right)
  +
  \sum_j
  \left| p_j \right|^2
  \log
  \left| p_j \right|^2
\end{equation}
%where $S_A$ is the von Neumann entropy~\eqref{von_Neumann_entropy}.
We shall compute the topological entanglement
entropy~\eqref{topological_entanglement} of the non-Abelian
quasi-particle states in the following.

We  explain how to compute the topological entanglement 
based on the skein theory.
In our formulation, the state  is
schematically written as
\begin{equation}
  \left| \Psi \right\rangle
  =
  \mbox{
    \raisebox{-.3cm}{
      \includegraphics[scale=0.8]{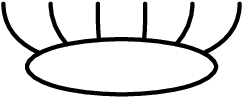}
    }
  }
\end{equation}
where each end of arcs denotes quasi-particles at spatially different
points.
There exists a sum of trivalent graphs insides an ellipse, which
is a Wilson line describing how quasi-particles fuse to the vacuum sector.
Correspondingly the density matrix~\eqref{density_matrix} is depicted
as
\begin{equation}
  \rho
  =
  \mbox{
    \raisebox{-.8cm}{
      \includegraphics[scale=0.8]{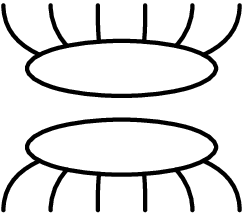}
    }
  }
\end{equation}
We now divide quasi-particles into  two groups $A$ and $B$.
We mean that the owner of each quasi-particle is Alice or Bob.
As the Alice's
reduced  density
matrix $\rho_A$~\eqref{reduced_density} is given by taking
traces in Bob's space, it is depicted as follows by connecting ends
of Bob's arcs;
\begin{equation}
  \label{depict_rho_A}
  \rho_A=
  \mbox{
    \raisebox{-11mm}{
      \includegraphics[scale=0.8]{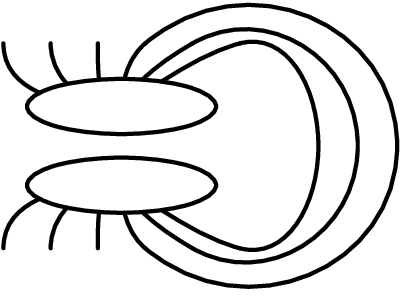}
    }
  }
  =
  \mbox{
    \raisebox{-8mm}{
      \includegraphics[scale=0.8]{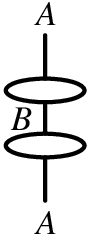}
    }
  }
\end{equation}
where in the last expression, $A$ and $B$ denote quasi-particles
belonging to Alice and Bob respectively, not spins of quasi-particles.
Applying~\eqref{entropy_replica},
the von Neumann entropy~\eqref{von_Neumann_entropy} is  computed
by connecting $n$ copies of
the reduced density matrix $\rho_A$;
\begin{equation}
  \label{depict_S_A}
  S_A
  = - \lim_{n\to 1 } \frac{\partial}{\partial n} \,
  \mbox{
    \raisebox{-10mm}{
      \includegraphics[scale=0.8]{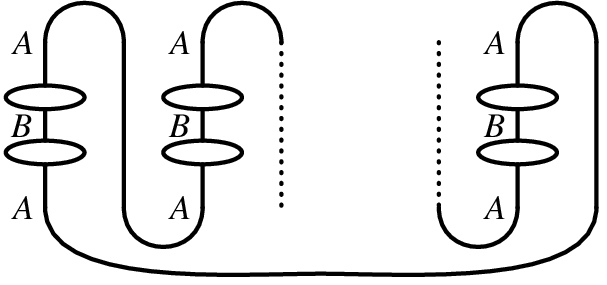}
    }
  }
\end{equation}
Above trivalent graphs are computable  by the skein theory, and we
can obtain explicitly  the topological 
entanglement entropy from~\eqref{topological_entanglement}.

%%%%%%%%%%%%%%%%%%%%%%%%
\section{Quasi-particles with Spin $1/2$}
\label{sec:spin-half}

\begin{figure}[tbhp]
  \centering
  \includegraphics[scale=0.6]{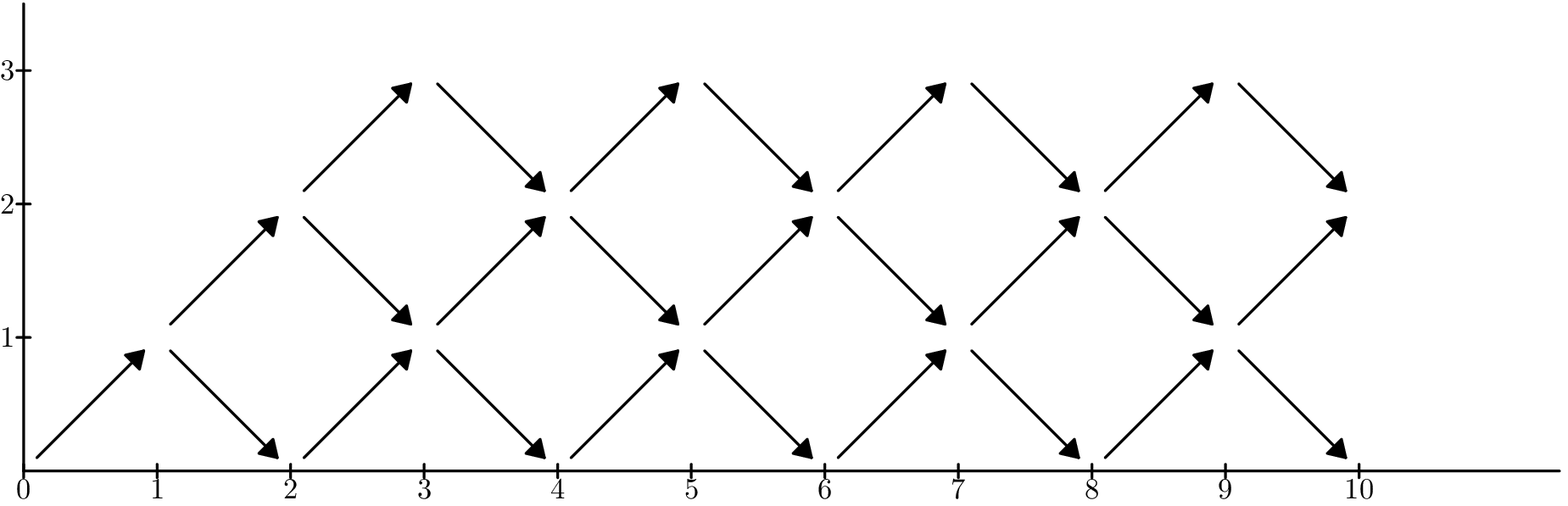}
  \caption{Bratteli diagram for spin-$1/2$ for $SU(2)_3$ theory.}
  \label{fig:Bratteli_spin-half}
\end{figure}

We study the braiding operations on the correlation function of
many-quasi-particles with  spin-$1/2$.
The Bratteli diagram of  spin-$1/2$ quasi-particles is depicted in
Fig.~\ref{fig:Bratteli_spin-half}.
One sees that the number of quasi-particles must be even so that the
correlation function is non-vanishing.
In the case of 2-quasi-particle state,  we have only a unique path
which starts from $(0,0)$ and ends at $(2,0)$ in the Bratteli diagram
in Fig.~\ref{fig:Bratteli_spin-half}, which corresponds to the
state~\eqref{normalized_qp}.
Generally we have several paths from $(0,0)$ to $(2\,n ,0)$.
The
number of paths 
denotes that of  the fusion channels, and it is the
dimension of the Hilbert space of $2 \, n$-quasi-particle states.
We see that,
for the $SU(2)_2$ theory,
it
%the number of path from $(0,0)$ to $(2 \, n , 0)$
is
$2^{n-1}$, which denotes  the dimension studied in
Ref.~\citenum{NayaWilc96a}.

We shall study the action of the braid operators on the
$2 \, n $-quasi-particle states.
The non-Abelian property of quasi-particles is that we have the
nontrivial representation of the braid operators.
Hereafter
$\sigma_i$ means the braid operator acting on the $i$-th and $i+1$-th
particles as
\begin{equation}
  \sigma_i =
  \mbox{
    \raisebox{-.6cm}{
      \includegraphics[scale=0.8]{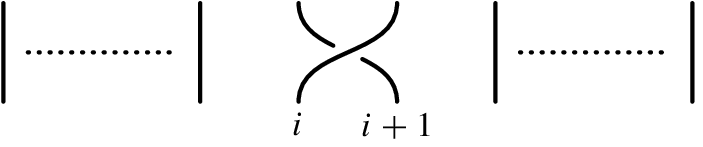}
    }
  }
\end{equation}
Here $i$ and $i+1$ mean the $i$-th and $i+1$-th positions from the
left respectively, and these should not be confused with spins of
quasi-particles.
The operators $\sigma_i$ satisfy the Artin braid relation
\begin{equation}
  \label{Artin_braid}
  \begin{aligned}
    \mbox{
      \raisebox{-1.4cm}{
        \includegraphics[scale=0.8]{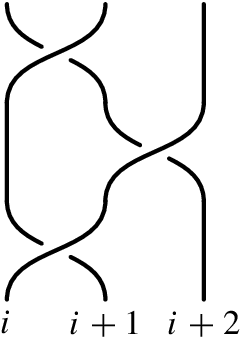}
      }
    }
    & =
    \mbox{
      \raisebox{-1.4cm}{
        \includegraphics[scale=0.8]{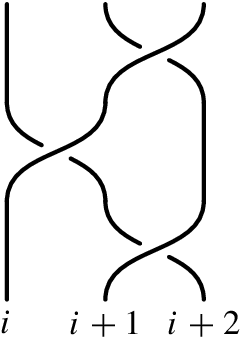}
      }
    }
    \\[2mm]
    \sigma_i \, \sigma_{i+1} \, \sigma_i
    & =
    \sigma_{i+1} \, \sigma_{i} \, \sigma_{i+1}
  \end{aligned}
\end{equation}
Easy is to see that
\begin{equation*}
  \sigma_i \, \sigma_j = \sigma_j \, \sigma_i
\end{equation*}
for $|i-j| \geq 2$.
We also use the twist $\theta_i$;
\begin{equation}
  \label{theta_twist}
  \theta_i =
  \mbox{
    \raisebox{-.6cm}{
      \includegraphics[scale=0.8]{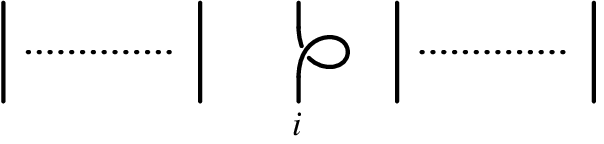}
    }
  }
\end{equation}

%%%%
\subsection{$4$-Quasi-Particle States}

We study the braiding effects in
the 4-quasi-particle states in the $SU(2)_{K \geq 2}$ theory.
A dimension of the Hilbert space of 4 quasi-particles is two, and we may
choose unnormalized bases as
\begin{equation*}
  \begin{gathered}
    |\widetilde{0} \rangle 
    =
    \mbox{
      \raisebox{-.1cm}{
        \includegraphics[scale=0.7]{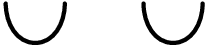}
      }
    }
    \\[2mm]
    |\widetilde{1} \rangle 
    =
    \mbox{
      \raisebox{-.3cm}{
        \includegraphics[scale=0.7]{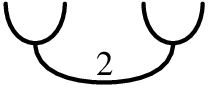}
      }
    }
  \end{gathered}
\end{equation*}
Correspondingly dual bases $\langle \widetilde{0}|$
and $\langle \widetilde{1}|$ are defined
upside down.
The norms of these diagrams are computed as
\begin{align*}
  \langle \widetilde{0} | \widetilde{0} \rangle
  & =
  \mbox{
    \raisebox{-.16cm}{
      \includegraphics[scale=0.7]{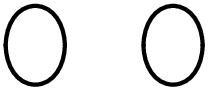}
    }}
  = d^2 
  \\[2mm]
  \langle \widetilde{1} | \widetilde{1} \rangle
  & =
  \mbox{
    \raisebox{-.6cm}{
      \includegraphics[scale=0.7]{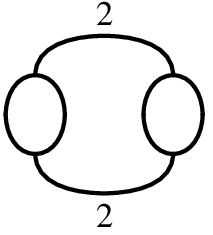}
    }}
  =
  \mbox{
    \raisebox{-.4cm}{
      \includegraphics[scale=0.7]{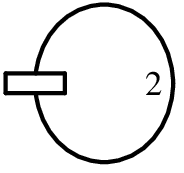}
    }}
  = \Delta_2
\end{align*}
where we have used
\begin{equation}
  \label{identity_proj_1}
  \mbox{
    \raisebox{-6mm}{
      \includegraphics[scale=0.7]{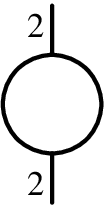}
    }}
  =
  \mbox{
    \raisebox{-6mm}{
      \includegraphics[scale=0.7]{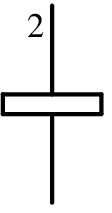}
    }
  }
\end{equation}
We see   that
they have positive norms,
$d^2>0$ and $\Delta_2 >0$,
under a parametrization~\eqref{A_set_i_times}.
It is noted that,
when $K=1$, we have $\Delta_2=0$, and that
$\left|\widetilde{1}\right\rangle$ becomes a null state
due to that it does not satisfy the admissible
condition~\eqref{admissible}.
We also see that
\begin{equation*}
  \langle \widetilde{0} | \widetilde{1} \rangle
  =
  \mbox{
    \raisebox{-.4cm}{
      \includegraphics[scale=0.7]{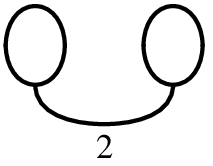}
    }}
  = 0
\end{equation*}
by making use of the following identity, which is a special case
of~\eqref{projection_zero};
\begin{equation}
  \label{identity_proj_2}
  \mbox{
    \raisebox{-4mm}{
      \includegraphics[scale=0.7]{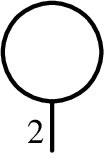}
    }}
  = 0
\end{equation}
Thus we indeed have two normalized physical states
\begin{equation}
  \label{normalized_4_qp_half}
  \begin{gathered}
    |0 \rangle 
    =
    \frac{1}{d} \,
    \mbox{
      \raisebox{-.1cm}{
        \includegraphics[scale=0.7]{qubit.1.eps}
      }
    }
    \\[2mm]
    |1 \rangle 
    =
    \frac{1}{\sqrt{d^2-1}} \,
    \mbox{
      \raisebox{-.3cm}{
        \includegraphics[scale=0.7]{qubit.2.eps}
      }
    }
  \end{gathered}
\end{equation}
which are orthonormal;
for $i.j \in \{0,1\}$ we have
\begin{align}
  \langle i | j \rangle = \delta_{i,j}
%   &
%   \langle 0 | 0 \rangle
%   =
%   \langle 1 | 1 \rangle
%   =
%   1
%   &
%   \langle 1 | 0 \rangle =
%   \langle 0 | 1 \rangle = 0
\end{align}

We study actions of braid operators on states
$\{ |0\rangle , | 1\rangle\}$.
Here we use the $F$-matrix defined by
\begin{equation}
  \label{base_F_qubit}
  \begin{pmatrix}
    \mbox{
      \raisebox{-.4cm}{
        \includegraphics[scale=0.6]{skein2.1.eps}
      }
    }
    \\[2mm]
    \mbox{
      \raisebox{-.5cm}{
        \includegraphics[scale=0.6]{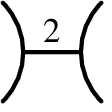}
      }
    }
  \end{pmatrix}
  =
  \mathbf{F}_{11}^{11} \,
  \begin{pmatrix}
    \mbox{
      \raisebox{-.4cm}{
        \includegraphics[scale=0.6]{skein3.1.eps}
      }
    }
    \\[2mm]
    \mbox{
      \raisebox{-.5cm}{
        \includegraphics[scale=0.6]{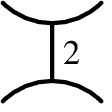}
      }
    }
  \end{pmatrix}
\end{equation}
We see that
\begin{equation}
  \mathbf{F}_{11}^{11}
  =
  \begin{pmatrix}
    \frac{1}{d} & 1 \\
    1-\frac{1}{d^2} & - \frac{1}{d}
  \end{pmatrix}
\end{equation}
which can be computed by use of a recursion relation of the Jones--Wenzl
idempotent~\eqref{Jones-Wenzl}.
Actions of $\sigma_1$ and $\sigma_3$ can be computed easily by use
of~\eqref{twist_simple}.
For instance, we have
\begin{align*}
  \sigma_1 \, 
  \mbox{
    \raisebox{-.1cm}{
      \includegraphics[scale=0.7]{qubit.1.eps}
    }
  }
  & =
  \mbox{
    \raisebox{-.1cm}{
      \includegraphics[scale=0.7]{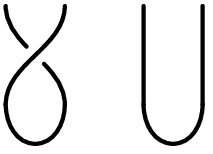}
    }
  }
  = - A^{-3} \,
  \mbox{
    \raisebox{-.1cm}{
      \includegraphics[scale=0.7]{qubit.1.eps}
    }
  }
\end{align*}
To get an action of $\sigma_2$, the $F$-matrix~\eqref{base_F_qubit}
will be useful.
We have
\begin{align*}
  \sigma_2 \,
  \begin{pmatrix}
    \mbox{
      \raisebox{-.1cm}{
        \includegraphics[scale=0.6]{qubit.1.eps}
      }
    }
    \\[2mm]
    \mbox{
      \raisebox{-.1cm}{
        \includegraphics[scale=0.6]{qubit.2.eps}
      }
    }
  \end{pmatrix}
  & =
  \mathbf{F}_{11}^{11} \,
  \begin{pmatrix}
    \mbox{
      \raisebox{-.3cm}{
        \includegraphics[scale=0.6]{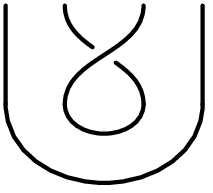}
      }
    }
    \\[2mm]
    \mbox{
      \raisebox{-.3cm}{
        \includegraphics[scale=0.6]{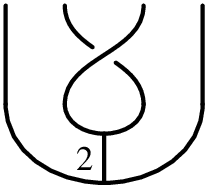}
      }
    }
  \end{pmatrix}
  \\
  & =
  \mathbf{F}_{11}^{11} \,
  \begin{pmatrix}
    -A^3      & 0
    \\[2mm]
    0 & A
  \end{pmatrix}
  \,
  \begin{pmatrix}
    \mbox{
      \raisebox{-.3cm}{
        \includegraphics[scale=0.6]{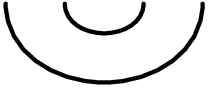}
      }
    }
    \\[2mm]
    \mbox{
      \raisebox{-.3cm}{
        \includegraphics[scale=0.6]{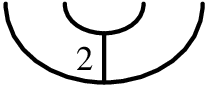}
      }
    }
  \end{pmatrix}
  \\
  & =
  \mathbf{F}_{11}^{11} \,
  \begin{pmatrix}
    -A^3      & 0
    \\[2mm]
    0 & A
  \end{pmatrix}
  \,
  \left(\mathbf{F}_{11}^{11}\right)^{-1} \,
  \begin{pmatrix}
    \mbox{
      \raisebox{-.3cm}{
        \includegraphics[scale=0.6]{qubit.1.eps}
      }
    }
    \\[2mm]
    \mbox{
      \raisebox{-.3cm}{
        \includegraphics[scale=0.6]{qubit.2.eps}
      }
    }
  \end{pmatrix}
\end{align*}
Actions of       other operators can be computed in the same way.
In summary,
the representation of the braid operators on
the space spanned by
$\left\{ | 0 \rangle , | 1 \rangle \right\}$~\eqref{normalized_4_qp_half}
are as follows;
\begin{equation}
  \begin{gathered}
%    R_{12} = R_{34}=
    \rho(\sigma_1) =
    \rho(\sigma_3) =
    \begin{pmatrix}
      -A^{-3} & \\
      & A
    \end{pmatrix}
    \\[2mm]
% 
%    R_{23} =
    \rho(\sigma_2) =
    \begin{pmatrix}
      -\frac{A^3}{d} & \frac{\sqrt{d^2-1}}{A \, d}
      \\[2mm]
      \frac{\sqrt{d^2-1}}{A \, d} 
      & \frac{1}{A^5 \, d}
    \end{pmatrix}
  \end{gathered}
  \label{R_matrix_2d}
\end{equation}
which are unitary under~\eqref{A_set_i_times}.
We can check directly that these representations satisfy the
Artin braid
relation~\eqref{Artin_braid}, and that
\begin{equation*}
  \rho \left(
    \sigma_1 \,    \sigma_2 \,    \sigma_3^{~2} \,
   \sigma_2 \,    \sigma_1 \,
  \right)
  =
  A^{-6} \cdot \mathbf{1}
\end{equation*}

Computation of the representation of the twist~\eqref{theta_twist} is
straightforward.
As all quasi-particles are spin-$1/2$, the representation of
$\theta_i$ is same for all $i$, and we have
\begin{equation}
  \label{twist_spin-half}
  \rho(\theta)
  =-A^3
\end{equation}

It is remarked that these representations are
computed in Ref.~\citenum{ArdonSchou07a} from an explicit form of
wave functions
as the  4-point correlation functions of the WZW model~\cite{KniZam84}.

\subsection{$6$-Quasi-Particle States}
We next study 6-quasi-particle states.
We restrict to the case $K>2$.
The $SU(2)_2$ case will be discussed separately in the later section.
As can be seen from the Bratteli diagram in
Fig.~\ref{fig:Bratteli_spin-half},
the dimension of the Hilbert spaces is $5$.
We set
\begin{equation}
  \label{base_6-particle}
  \begin{aligned}
    |00\rangle
    & =
    \frac{1}{d \, \sqrt{d}} \,
    \mbox{
      \raisebox{-.1cm}{
        \includegraphics[scale=0.7]{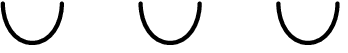}
      }
    }
    \\
    |01\rangle
    & =
    \frac{1}{\sqrt{d \, \left( d^2-1 \right)}} \,
    \mbox{
      \raisebox{-.3cm}{
        \includegraphics[scale=0.7]{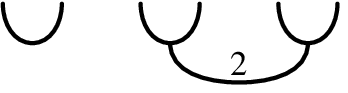}
      }
    }
    \\
    |10\rangle
    & =
    \frac{1}{\sqrt{d \, \left( d^2-1 \right)}} \,
    \mbox{
      \raisebox{-.3cm}{
        \includegraphics[scale=0.7]{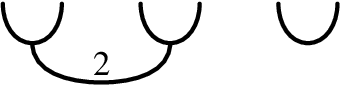}
      }
    }
    \\
    |11\rangle
    & =
    \frac{1}{\sqrt{d \, \left( d^2-1 \right)}} \,
    \mbox{
      \raisebox{-.3cm}{
        \includegraphics[scale=0.7]{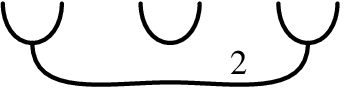}
      }
    }
    \\
    |C\rangle
    & =
    \sqrt{
      \frac{d}{\left( d^2 -1 \right) \,
        \left( d^2 -2 \right)}
    } \,
    \mbox{
      \raisebox{-.3cm}{
        \includegraphics[scale=0.7]{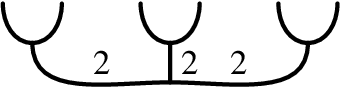}
      }
    }
  \end{aligned}
\end{equation}
We have defined  these states based on Ref.~\citenum{LGeorg06b}.
See that~\eqref{A_set_i_times} proves positivity of norm of each
diagram,
and
by the same computation with a case of 4-quasi-particle, we see that
these are orthonormal bases
\begin{equation*}
  \langle i | j \rangle = \delta_{i,j}
\end{equation*}
where
$i,j \in \{ 00,01,10,11,C \}$.

We study the action of the braid operators
$\left\{\sigma_1 , \dots, \sigma_5 \right\}$.
Actions of $\sigma_1$, $\sigma_3$, and $\sigma_5$ are easily computed
by use of~\eqref{twist_formula}.
Actions of $\sigma_2$ on $|00\rangle$ and $|10\rangle$,
and of $\sigma_4$ on $|00\rangle$ and $|01\rangle$,
can be given in
the same method with 4-quasi-particles.
For the action of others, we may use
\begin{equation}
  \label{F_1212}
  \begin{pmatrix}
    \mbox{
      \raisebox{-.2cm}{
        \includegraphics[scale=0.6]{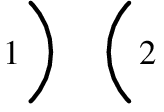}
      }
    }
    \\[2mm]
    \mbox{
      \raisebox{-.2cm}{
        \includegraphics[scale=0.6]{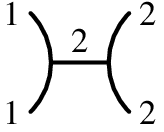}
      }
    }
  \end{pmatrix}
  =
  \mathbf{F}_{12}^{12} \,
  \begin{pmatrix}
    \mbox{
      \raisebox{-.2cm}{
        \includegraphics[scale=0.6]{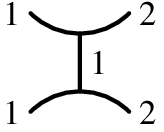}
      }
    }
    \\[2mm]
    \mbox{
      \raisebox{-.2cm}{
        \includegraphics[scale=0.6]{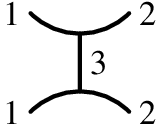}
      }
    }
  \end{pmatrix}
\end{equation}
Here the $F$-matrix is given by
\begin{equation}
  \mathbf{F}_{12}^{12}
  =
  \begin{pmatrix}
    \frac{d}{d^2-1} & 1
    \\[2mm]
    1- \frac{1}{d^2-1} & -\frac{1}{d}
  \end{pmatrix}
\end{equation}
which is derived by the recursion relation of the Jones-Wenzl
idempotent~\eqref{Jones-Wenzl}.
Then, for example, we compute as follows;
\begin{align*}
  \sigma_2 \,
  \mbox{
    \raisebox{-.3cm}{
      \includegraphics[scale=0.6]{qu6.3.eps}
    }
  }
  & =
  \mbox{
    \raisebox{-.5cm}{
      \includegraphics[scale=0.6]{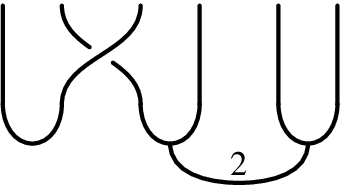}
    }
  }
  \\
  & =
  \left( \mathbf{F}_{11}^{11} \right)_{11} \,
  \mbox{
    \raisebox{-.5cm}{
      \includegraphics[scale=0.6]{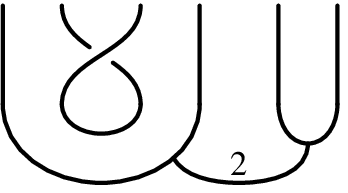}
    }
  }
  +
  \left( \mathbf{F}_{11}^{11} \right)_{11} \,
  \mbox{
    \raisebox{-.5cm}{
      \includegraphics[scale=0.6]{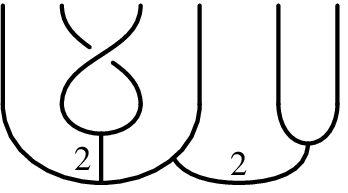}
    }
  }
  \\
  & =
  \left( \mathbf{F}_{11}^{11} \right)_{11} \,
  \left( - A^{-3} \right) \,
  \mbox{
    \raisebox{-.3cm}{
      \includegraphics[scale=0.6]{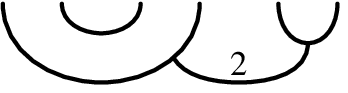}
    }
  }
  +
  \left( \mathbf{F}_{11}^{11} \right)_{11} \,
  A \,
  \mbox{
    \raisebox{-.3cm}{
      \includegraphics[scale=0.6]{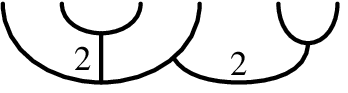}
    }
  }
  \\
  & =
  \left(
    \mathbf{F}_{11}^{11} \,
    \begin{pmatrix}
      -A^{-3} & 0
      \\
      0 & A
    \end{pmatrix} \,
    \left( \mathbf{F}_{11}^{11} \right)^{-1}
  \right)_{11} \,
  \mbox{
    \raisebox{-.3cm}{
      \includegraphics[scale=0.6]{qu6.3.eps}
    }
  }
  \\
  & \qquad
  +
  \left(
    \mathbf{F}_{11}^{11} \,
    \begin{pmatrix}
      -A^{-3} & 0
      \\
      0 & A
    \end{pmatrix} \,
    \left( \mathbf{F}_{11}^{11} \right)^{-1}
  \right)_{12} \,
  \mbox{
    \raisebox{-.3cm}{
      \includegraphics[scale=0.6]{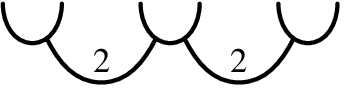}
    }
  }
\end{align*}
In the last expression, we further apply~\eqref{F_1212} to get
\begin{equation*}
  \mbox{
    \raisebox{-.3cm}{
      \includegraphics[scale=0.6]{qu6.36.eps}
    }
  }
  =
  \left( (\mathbf{F}_{12}^{12})^{-1} \right)_{11} \,
  \mbox{
    \raisebox{-.3cm}{
      \includegraphics[scale=0.6]{qu6.4.eps}
    }
  }
  +
  \left( (\mathbf{F}_{12}^{12})^{-1} \right)_{12} \,
  \mbox{
    \raisebox{-.3cm}{
      \includegraphics[scale=0.6]{qu6.5.eps}
    }
  }
\end{equation*}
Other actions can be computed in the similar methods.

To summarize,
the representation of the braid operators on the 5-dimensional space
$\left\{
  |00 \rangle,   |01 \rangle,   |10 \rangle,
  |11 \rangle,   |C \rangle
\right\}$
are given by
\begin{equation}
  \begin{gathered}
%    R_{12} =
    \rho(\sigma_1) =
    \diag \left( -A^{-3}, -A^{-3},  A , A , A \right)
    \\[2mm]
%    R_{34}=
% \begin{align}
%   R_{12} & =
%   \begin{pmatrix}
%     -A^{-3} & & & & \\
%     & -A^{-3} & & & \\
%     & & A & & \\
%     & & & A & \\
%     & & & & A
%   \end{pmatrix}
%   \\
% %
%   R_{34} & =
%   \begin{pmatrix}
%     -A^{-3} & & & & \\
%     & A & & & \\
%     & & A & & \\
%     & & & -A^{-3}  & \\
%     & & & & A
%   \end{pmatrix}
%   \\
% %
%   R_{56} & =
%   \begin{pmatrix}
%     -A^{-3} & & & & \\
%     & A & & & \\
%     & & -A^{-3} & & \\
%     & & & A & \\
%     & & & & A
%   \end{pmatrix}
% \end{align}
%  \\
%    R_{23} =
    \rho(\sigma_2) =
    \begin{pmatrix}
      -\frac{A^3}{d} & 0 & \frac{\sqrt{d^2-1}}{A \, d} & 0 & 0
      \\[2mm]
      0 &
      -\frac{A^3}{d}
% A + \frac{1}{A \, d}
      &
      0 & \frac{1}{A \, d} &
      \frac{\sqrt{d^2-2}}{A \, d}
      \\[2mm]
      \frac{\sqrt{d^2-1}}{A \, d}   & 0 &
      \frac{1}{A^5  \, d} & 0 &  0
      \\[2mm]
      0 & \frac{1}{A \, d} & 0 &
      -\frac{A^3}{d}
%    A + \frac{1}{A \, d}
      & 
      \frac{\sqrt{d^2-2}}{A \, d}
      \\[2mm]
      0 &  \frac{\sqrt{d^2-2}}{A \, d} & 0 &
      \frac{\sqrt{d^2-2}}{A \, d}& 
      A+\frac{d^2-2}{A \,d }
    \end{pmatrix}
    \\[2mm]
%%%%
    \rho(\sigma_3) =
    \diag \left(
      -A^{-3}, A, A, -A^{-3}, A
    \right)
    \\[2mm]
    \rho(\sigma_4) =
%    R_{45} =
    \begin{pmatrix}
      -\frac{A^3}{d} &  \frac{\sqrt{d^2-1}}{A \,d} & 0 & 0 & 0
      \\[2mm]
      \frac{\sqrt{d^2-1}}{A \, d } &
      \frac{1}{A^5  \, d} & 0 &  0  & 0
      \\[2mm]
      0 & 0 &
      -\frac{A^3}{d}
%    A + \frac{1}{A \, d}
      & \frac{1}{A \, d} &
      \frac{\sqrt{d^2-2}}{A \,d }
      \\[2mm]
      0 & 0 &
      \frac{1}{A \, d} &
      -\frac{A^3}{d}
%    A + \frac{1}{A \, d}
      &
      \frac{\sqrt{d^2-2}}{A \,d }
      \\[2mm]
      0 & 0 & 
      \frac{\sqrt{d^2-2}}{A \,d }  & 
      \frac{\sqrt{d^2-2}}{A \,d }  & 
      A+\frac{d^2-2}{A \,d} 
    \end{pmatrix}
    \\[2mm]
%%%%
%    R_{56} =
    \rho(\sigma_5) =
    \diag \left(
      -A^{-3}, A, -A^{-3}, A, A
    \right)
  \end{gathered}
  \label{braid_6_quasi}
\end{equation}
One can check that these satisfy~\eqref{Artin_braid}, and
\begin{equation*}
  \rho \left(
    \sigma_1 \,    \sigma_2 \,    \sigma_3 \, \sigma_4 \, \sigma_5\,
    \sigma_5 \, \sigma_4 \,
    \sigma_3 \,    \sigma_2 \,    \sigma_1 \,
  \right)
  =
  A^{-6} \cdot \mathbf{1}
\end{equation*}
Note that the representation of the twist~\eqref{theta_twist} is same
with before, and we have~\eqref{twist_spin-half}.
% \begin{equation}
%   \rho(\theta)
%   =-A^3
% \end{equation}

%%%%%
\subsection{Topological Entanglement Entropy of Spin-$1/2$
  Quasi-Particle States}

%%%
\subsubsection{2-Quasi-Particle State}
As the first simple example of explicit computation of entanglement
entropy,
we study the 2-quasi-particle
state~\eqref{normalized_qp}.
We have the state
$\left|\Psi \right\rangle
=
\left| \text{qp} \right\rangle$,
and 
the density matrix $\rho$ is depicted as
\begin{equation}
  \rho = \frac{1}{d} \,
  \mbox{
    \raisebox{-6mm}{
      \includegraphics[scale=0.8]{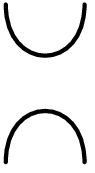}
    }
  }
\end{equation}
We assume that both Alice and Bob have one quasi-particle.
Alice (resp. Bob) has the left (resp. right) quasi-particle,
and we denote  it as
$A=\{ 1 \}$ and
$B=\{ 2\}$.
% We regard that two quasi-paricles correspond to two sub-regions $A$
% and $B$.
As
a  trace $\Tr_B$
means connecting arc ends of Bob's
quasi-particles~\eqref{depict_rho_A},
we have
\begin{equation}
  \rho_{\{1\}}= \Tr_{\{ 2 \}} \rho
  =
  \frac{1}{d} \,
  \mbox{
    \raisebox{-.7cm}{
      \includegraphics[scale=0.8]{twist.0.eps}
    }
  }
\end{equation}
Then from~\eqref{depict_S_A} 
we obtain the entanglement entropy as
\begin{equation}
  S_{\{ 1 \}} = 
  - \lim_{n\to 1 } \frac{\partial}{\partial n}
  \left(
    \frac{1}{d^n} \,
    \mbox{
      \raisebox{-3mm}{
        \includegraphics[scale=0.8]{circle.1.eps}
      }
    }
  \right)
  =
  \log d
\end{equation}
and the topological entanglement
entropy~\eqref{topological_entanglement} is identified with
\begin{equation}
  \label{qp_S_1_spin-12}
  S_{\{1 \}}^{\text{topo}}
  =
  S_{\{ 1 \} }
  =\log d  
\end{equation}
This result is due to
that Alice and Bob
are  intertwined by a
Wilson line of spin-$1/2$ 
quasi-particle  whose quantum dimension is
$d$~\eqref{quantum_dimension}.

%%%
\subsubsection{4-Quasi-Particle States}

We next consider  the 4-quasi-particle
states~\eqref{normalized_4_qp_half}, which spans two-dimensional space.
We take a  state as
\begin{equation}
  \label{general_4_qp_half}
  \left| \Psi \right\rangle
  = p_0 \, \left| 0 \right\rangle
  + p_1 \, \left| 1 \right\rangle  
\end{equation}
where
$\left|p_0\right|
+\left|p_1\right|^2=1$.
We first assume that the first and the second quasi-particles from the
left belong to Alice, and that the remaining   third and fourth
quasi-particles to  Bob;
we denote 
$A=\left\{1,2 \right\}$
and
$B=\left\{3,4 \right\}$.
Connecting ends of  Wilson lines of Bob's
quasi-particles as
in~\eqref{depict_rho_A}
and using~\eqref{identity_proj_1} and~\eqref{identity_proj_2},
Alice's reduced density matrix is schematically written as
\begin{equation}
  \rho_{\{1 , 2 \} }
  =
  \left|p_0\right|^2 \,
  \frac{1}{d} \,
  \mbox{
    \raisebox{-6mm}{
      \includegraphics[scale=0.8]{reduced.3.eps}
    }
  }
  +
  \left|p_1\right|^2
  \frac{1}{d^2 -1} \,
  \mbox{
    \raisebox{-6mm}{
      \includegraphics[scale=0.8]{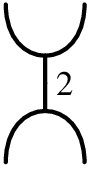}
    }
  }
\end{equation}
It is  straightforward to compute~\eqref{depict_S_A}, and we get
the von Neumann entropy as
\begin{align}
  S_{\{ 1, 2 \}}
  & =
  - \lim_{n\to 1} \frac{\partial}{\partial n} \,
  \left(
    \left(
      \frac{\left| p_0 \right|^2}{d}
    \right)^n \,
    \left(
      \mbox{
        \raisebox{-4mm}{
          \includegraphics[scale=0.8]{circle.1.eps}
        }
      }
    \right)^n
    +
    \left(
      \frac{\left| p_1 \right|^2}{d^2-1}
    \right)^n \,
    \mbox{
      \raisebox{-5mm}{
        \includegraphics[scale=0.7]{orthobase.3.eps}
      }}
    \right)
  \nonumber
  \\
  & =
  -
  \left|p_0\right|^2
  \log
  \left|p_0\right|^2
  -
  \left|p_1\right|^2
  \log \left(
    \frac{
      \left|p_1\right|^2
    }{
      d^2-1}
  \right)
  \label{entropy_4_half}
\end{align}
where we have used~\eqref{identity_proj_1}
and~\eqref{identity_proj_2}.
To extract a topological entropy,
we notice that the state~\eqref{general_4_qp_half} 
has the orthogonal bases 
$\left| \psi_{j} \right\rangle_{ \{1,2 \} }$
and
$\left| \phi_{j} \right\rangle_{ \{3,4 \} }$
as~\eqref{orthogonal_base} thanks to~\eqref{identity_proj_1}
and~\eqref{identity_proj_2}.
As a result,
the topological entanglement
entropy~\eqref{topological_entanglement} 
is identified with
\begin{equation}
  \label{topology_S_2_1}
  S_{\{1 , 2 \}}^{\text{topo}}=
  \left| p_1 \right|^2 \log
  \left( d^2 -1 \right)
\end{equation}
This result is interpreted as follows.
The state $|0\rangle$ is not topologically entangled between Alice and
Bob,
or
they are entangled by the vacuum sector with quantum dimension $d_0=1$.
Although, in the state $|1\rangle$ Alice's quasi-particles are
intertwined with Bob's through the Wilson line of quasi-particle with
spin-$1$ whose quantum dimension~\eqref{quantum_dimension}
is $d_2=\Delta_2=d^2 -1$.

% This result means that two groups  $A$ and $B$ are respectivley
% connected
% by spin-$0$
% and spin-$1$ for the state $\left|0\right\rangle$ and
% $\left|1\right\rangle$ whose quantum
% dimensions
% are
% $d_0=1$ and $d_2=\Delta_2=d^2-1$.

We make similar  computations for the 4-quasi-particle
states~\eqref{normalized_4_qp_half}
when Alice owns the second and the third particles while
Bob has remainder;
$A=\left\{2,3 \right\}$
and
$B=\left\{1,4 \right\}$.
Following~\eqref{depict_rho_A}
the reduced density matrix is given by
\begin{align}
  \rho_{\{ 2,3 \}}
  & =
  \frac{\left| p_0 \right|^2}{d^2} \,
  \mbox{
    \raisebox{-6mm}{
      \includegraphics[scale=0.8]{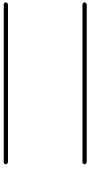}
    }
  }
  +
  \frac{p_0\, p_1^* + p_0^* \, p_1}{d \, \sqrt{d^2-1}} \,
  \mbox{
    \raisebox{-6mm}{
      \includegraphics[scale=0.8]{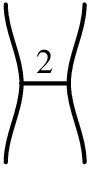}
    }
  }
  +
  \frac{\left|p_1\right|^2}{d^2 - 1} \,
  \mbox{
    \raisebox{-6mm}{
      \includegraphics[scale=0.8]{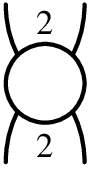}
    }
  }
  \nonumber
  \\
  & =
  \frac{\left| \widetilde{p_1} \right|^2}{d^2 -1} \,
%   \left|
%     \frac{p_0}{d} - \frac{p_1}{d \, \sqrt{d^2 -1}}
%   \right|^2 \,
  \mbox{
    \raisebox{-.8cm}{
      \includegraphics[scale=0.8]{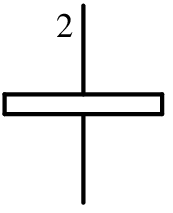}
    }
  }
  +
%   d \,
%   \left|
%     \frac{p_0}{d^2} + p_1 \, \frac{\sqrt{d^2-1}}{d^2}
%   \right|^2
  \frac{\left| \widetilde{p_0} \right|^2}{d}
  \,
  \mbox{
    \raisebox{-6mm}{
      \includegraphics[scale=0.8]{reduced.3.eps}
    }
  }
\end{align}
where
\begin{equation}
  \begin{aligned}
    \widetilde{p_0}
    & =
    \frac{p_0}{d} + p_1 \, \frac{\sqrt{d^2-1}}{d}
    \\[2mm]
    \widetilde{p_1}
    & =
    p_0 \,
    \frac{\sqrt{d^2-1}}{d}
    - \frac{p_1}{d}
  \end{aligned}
\end{equation}
Here, in the second equality, we have substituted a definition of
a  trivalent vertex~\eqref{trivalent} and have used the Jones--Wenzl
idempotent~\eqref{Jones-Wenzl}.
Note that
$\left| \widetilde{p_0} \right|^2
+\left| \widetilde{p_1} \right|^2=1$.
Using~\eqref{projection_zero} 
the entanglement entropy~\eqref{depict_S_A} is computed as
\begin{align}
  S_{\{ 2,3\}}
  & =
  - \lim_{n\to 1} \frac{\partial}{\partial n} \,
  \left(
    \left(
      \frac{\left| \widetilde{p_1} \right|^2}{d^2-1}
    \right)^n \,
    \mbox{
      \raisebox{-5mm}{
        \includegraphics[scale=0.7]{orthobase.3.eps}
      }}
    +
    \left(
      \frac{\left| \widetilde{p_0} \right|^2}{d}
    \right)^n \,
    \left(
      \mbox{
        \raisebox{-4mm}{
          \includegraphics[scale=0.8]{circle.1.eps}
        }
      }
    \right)^n
  \right)
  \nonumber
  \\
  & =
  -
  \left|\widetilde{p_1}\right|^2
  \log \left(
    \frac{
      \left| \widetilde{p_1}\right|^2
    }{
      d^2-1}
  \right)
  -
  \left|\widetilde{p_0}\right|^2
  \log
  \left|\widetilde{p_0}\right|^2
  \label{entropy_4_half_2}
\end{align}
To extract the topological entanglement entropy,
we should note that bases in~\eqref{normalized_4_qp_half} do not give
orthogonal bases,
$\left| \psi_{j} \right\rangle_{ \{2,3 \} }$
and
$\left| \phi_{j} \right\rangle_{ \{1,4 \} }$,
in Alice's and Bob's spaces~\eqref{orthogonal_base}.
To remedy it we rewrite the
state~\eqref{general_4_qp_half}  by use of the
$F$-matrix~\eqref{base_F_qubit}
as
\begin{equation}
  \label{general_4_qp_half_2}
  \left|\Psi \right\rangle
  =
  \widetilde{p_0} \, \frac{1}{d} \,
  \mbox{
    \raisebox{-.3cm}{
      \includegraphics[scale=0.8]{s2qubit.1.eps}
    }
  }
  +
  \widetilde{p_1} \, \frac{1}{\sqrt{d^2-1}} \,
  \mbox{
    \raisebox{-.3cm}{
      \includegraphics[scale=0.8]{s2qubit.2.eps}
    }
    }
\end{equation}
which has a form of~\eqref{orthogonal_base},
\emph{i.e.}, the states $\left|\psi_{j}\right\rangle_{ \{2,3 \} }$
and $\left|\phi_{j }\right\rangle_{ \{1,4 \} }$
are  orthonormal due to~\eqref{identity_proj_1} and~\eqref{identity_proj_2}
(see, \emph{e.g.},~\eqref{trivalent_theta}).
Thus
the topological entanglement entropy~\eqref{von_Neumann_entropy}
is given by
\begin{equation}
  \label{topology_S_2_2}
  S_{\{2,3 \}}^{\text{topo}}=
  \left| \widetilde{p_1} \right|^2 \log
  \left( d^2 -1 \right)
\end{equation}
When we look at the expression~\eqref{general_4_qp_half_2},
we see that Alice's and Bob's quasi-particles are
not topologically entangled with a probability $\left|\widetilde{p_0}\right|^2$
while they are intertwined with a probability
$\left|\widetilde{p_1}\right|^2$
by the Wilson line
of quasi-particle with spin-$1$ whose quantum dimension  is
$d_2=d^2-1$.
Thus the
expression~\eqref{general_4_qp_half_2} is comparable
with~\eqref{general_4_qp_half} replacing $p_a$ with $\widetilde{p_a}$.

% So
% results~\eqref{entropy_4_half} and~\eqref{entropy_4_half_2} supports
% our interpretation.

We leave it for readers to check that
\begin{equation}
  \label{topology_S_single_1}
  S_{\{1\}}^{\text{topo}}
  =
  S_{\{2,3,4\}}^{\text{topo}}
  =
  \log d
\end{equation}
which supports our interpretation because Alice and Bob are intertwined
by spin-$1/2$ quasi-particle
(consult computations in Section~\ref{sec:topological_entropy}).
In view of~\eqref{topology_S_2_1},~\eqref{topology_S_2_2},
and~\eqref{topology_S_single_1}, the topological entanglement entropy
depends on what quasi-particles Alice and Bob  have in
the state $\left|\Psi \right\rangle$.
Even if the state is same,
there exists a possibility that a different
quantum dimension appears depending on a quasi-particle which
intertwines Alice
and Bob when we change owners of quasi-particles.

% Later in this article
% we will give a general setup of the topological entanglement entropy
% for 4-quasi-particle states.

%%%%%%
%%%%%%%%
\subsection{Pfaffian State}

The Moore--Read Pfaffian state is described by the $SU(2)_2$ CS
theory,
and we have
$A=\I \, \E^{\pi \I/8}$ and
$d=\sqrt{2}$.
We then have a representation of the twist $\theta_i$~\eqref{theta_twist} as
\begin{equation}
  \rho(\theta) = - \E^{-\pi \I/8}
\end{equation}
The braid matrices~\eqref{R_matrix_2d} for 4-quasi-particle
states~\eqref{normalized_4_qp_half}
become
\begin{equation}
  \label{braiding_4-qp_Pfaffian}
  \begin{gathered}
    \rho(\sigma_1) =
    \rho(\sigma_3) =
%    R_{12} = R_{34} =
    \E^{\pi \I/8} \,
    \begin{pmatrix}
      -1 &
      \\
      & \I
    \end{pmatrix}
    \\
%    R_{23} =
    \rho(\sigma_2) =
    -
    \frac{
      \E^{-\frac{1}{8} \pi \I}
    }{\sqrt{2}} \,
    \begin{pmatrix}
      1 & \I \\
      \I & 1
    \end{pmatrix}
  \end{gathered}
\end{equation}

For 6-quasi-particle, the state $|C \rangle$
amongst~\eqref{base_6-particle}
is not admissible due
to~\eqref{admissible};
we can see that
norm of the diagram
for $\left| C\right\rangle$
% in~\eqref{base_6-particle}
reduces to zero, and it
is unphysical, null state.
Then from~\eqref{braid_6_quasi} the braiding matrices on space spanned
by
bases
$\{
|00\rangle,
|01\rangle,
|10\rangle,
|11\rangle
\}$  in~\eqref{base_6-particle} are given by
\begin{equation}
  \begin{gathered}
    \rho(\sigma_1) =
%    R_{12} =
    \E^{\pi \I/8}
    \diag \left(
      -1 , -1 , \I, \I
    \right)
    \\[2mm]
    \rho(\sigma_2) =
%    R_{23}=
    - \frac{\E^{-\frac{1}{8} \pi \I}}{\sqrt{2}} \,
    \begin{pmatrix}
      1 & 0 & \I & 0 \\
      0 & 1 & 0 & \I \\
      \I & 0 & 1 & 0 \\
      0 & \I & 0 & 1
    \end{pmatrix}
    \\[2mm]
    \rho(\sigma_3) =
%    R_{34} =
    \E^{\pi \I/8}
    \diag \left(
      -1, \I, \I, -1
    \right)
    \\[2mm]
    \rho(\sigma_4) =
%    R_{45} =
    -
    \frac{\E^{- \frac{1}{8} \pi \I}}{\sqrt{2}} \,
    \begin{pmatrix}
      1 & \I & 0 & 0 \\
      \I & 1 & 0 & 0 \\
      0 & 0 & 1  & \I \\
      0 & 0 & \I & 1
    \end{pmatrix}
    \\[2mm]
    \rho(\sigma_5) =
%    R_{56} =
    \E^{\pi \I/8}
    \diag \left(
      -1 , \I , -1 , \I
    \right)
  \end{gathered}
\end{equation}
These representations are studied in Ref.~\citenum{LGeorg06a}.
Also see Ref.~\citenum{DAIvano01a} where studied are
the braiding matrices of vortices
in a $p$-wave superconductor.

Concerning the topological entanglement entropy,
we  see that the quantum dimensions~\eqref{quantum_dimension} 
are given by
$d_1= \sqrt{2}$ and $d_2 = 1$, and that
only spin-$1/2$
quasi-particles contribute to the topological entanglement entropy.
Our results,~\eqref{topology_S_2_1} and~\eqref{topology_S_2_2},
indicate that
the topological entanglement entropy
vanishes in the 4-quasi-particle states~\eqref{general_4_qp_half}
when both Alice and Bob have two quasi-particles.
Note that it does not vanish when Alice has one 
or three quasi-particles
(see~\eqref{topology_S_single_1}).

%%%%%%%%%%%%%%%%%%%5
\section{Quasi-particles with Spin-$1$}
\label{sec:spin-1}

\begin{figure}[tbhp]
  \centering
  \includegraphics[scale=0.6]{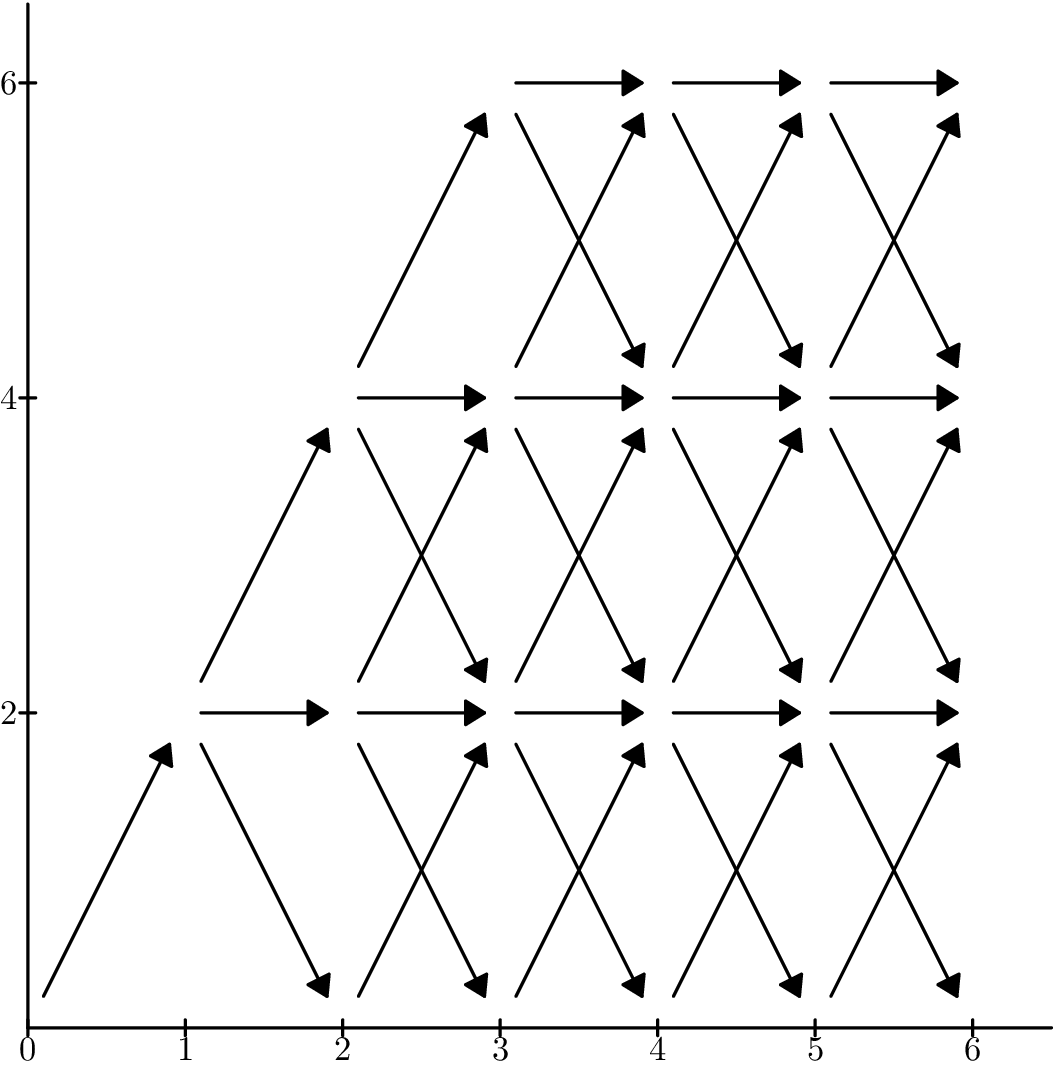}
  \caption{Bratteli diagram for spin-$1$ for $SU(2)_7$ theory}
  \label{fig:Bratteli_spin-1}
\end{figure}

We study the braid group operations on the correlation functions of
quasi-particles with spin-$1$.
The fusion rule of spin-$1$ gives the Bratteli diagram as in
Fig.~\ref{fig:Bratteli_spin-1}.
As in the previous section, the number of paths from $(0,0)$ to
$(n,0)$ is the dimension of the conformal block of
$n$-quasi-particles.
See  that,
different from the spin-$1/2$ case, we have the non-vanishing
correlation function even when $n$ is odd.
Both the dimension of 2- and 3-quasi-particle states  is one, and
the normalized 2-quasi-particle and 3-quasi-particle states are
respectively given by
\begin{gather}
  \label{qp_2_spin-1}
  \left| \text{qp}_2
  \right\rangle
  =
  \frac{1}{\sqrt{d^2-1}} \,
  \mbox{
    \raisebox{-.4cm}{
      \includegraphics[scale=0.8]{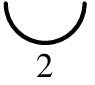}
    }
  }
  \\[2mm]
  \label{qp_3_spin-1}
  \left| \text{qp}_3
  \right\rangle
  =
  \sqrt{
    \frac{d}{
      \left( d^2 -1 \right) \,
      \left( d^2 - 2 \right)
    }
  }
  \mbox{
    \raisebox{-.3cm}{
      \includegraphics[scale=0.8]{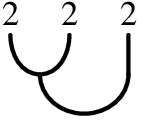}
    }
  }
\end{gather}
In $\left| \text{qp}_2 \right\rangle$
two endpoints of arc labelled $2$ indicates that there are
two quasi-particles
with spin-$1$ in spatially different points.
Correspondingly in $\left| \text{qp}_3 \right\rangle$
three ends of arcs denote that
there are three quasi-particles in spatially different points.

%%%%
\subsection{4-Quasi-Particles States}
We first assume
$K>3$.
The $SU(2)_3$ theory will be discussed later.
A dimension of the Hilbert space of 4-quasi-particles with spin-1 is
three
as can be read from the Bratteli diagram in
Fig.~\ref{fig:Bratteli_spin-1}, and
we set
\begin{equation}
  \label{1-qubit_spin_1}
  \begin{gathered}
    |0 \rangle 
    =
    \frac{1}{d^2-1} \,
    \mbox{
      \raisebox{-.2cm}{
        \includegraphics[scale=0.7]{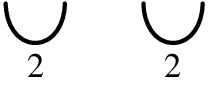}
      }
    }
    \\
    |1 \rangle 
    =
    \frac{d}{(d^2-2) \, \sqrt{d^2-1}} \,
    \mbox{
      \raisebox{-.3cm}{
        \includegraphics[scale=0.7]{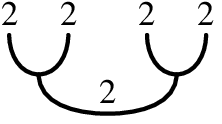}
      }
    }
    \\
    |2 \rangle 
    =
    \frac{1}{\sqrt{d^4- 3 \, d^2 +1}} \,
    \mbox{
      \raisebox{-.3cm}{
        \includegraphics[scale=0.7]{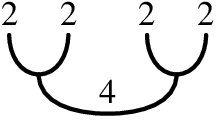}
      }
    }
  \end{gathered}
\end{equation}
A parameter~\eqref{A_set_i_times} supports positivity of norm of each
diagram, and
the recursion relation~\eqref{Jones-Wenzl} of
the Jones--Wenzl idempotent proves the orthonormality;
\begin{equation*}
  \langle i | j \rangle = \delta_{i,j}
\end{equation*}

The $F$-matrix which we use in this section is
\begin{equation}
  \label{F_matrix_22_22}
  \begin{pmatrix}
    \mbox{
      \raisebox{-.2cm}{
        \includegraphics[scale=0.6]{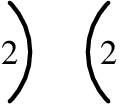}
      }
    }
    \\[2mm]
    \mbox{
      \raisebox{-.2cm}{
        \includegraphics[scale=0.6]{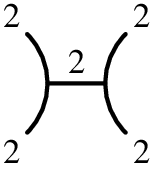}
      }
    }
    \\[2mm]
    \mbox{
      \raisebox{-.2cm}{
        \includegraphics[scale=0.6]{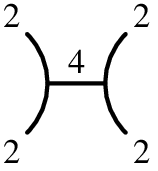}
      }
    }
  \end{pmatrix}
  =
  \mathbf{F}_{22}^{22} \,
  \begin{pmatrix}
    \mbox{
      \raisebox{-.2cm}{
        \includegraphics[scale=0.6]{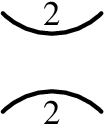}
      }
    }
    \\[2mm]
    \mbox{
      \raisebox{-.2cm}{
        \includegraphics[scale=0.6]{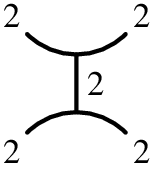}
      }
    }
    \\[2mm]
    \mbox{
      \raisebox{-.2cm}{
        \includegraphics[scale=0.6]{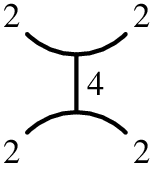}
      }
    }
  \end{pmatrix}
\end{equation}
where
\begin{equation}
  \mathbf{F}_{22}^{22}
  =
  \begin{pmatrix}
    \frac{1}{d^2-1} & \frac{d}{d^2-2} & 1
    \\[2mm]
    \frac{d^2-2}{d \, (d^2-1)} & 1 - \frac{1}{d^2-2} &
    -\frac{1}{d}
    \\[2mm]
    1-\frac{d^2}{(d^2 - 1)^2} &
    -\frac{d \, (d^4 - 3 \, d^2 +1)}{(d^2-1) \, (d^2 -2)^2}
    &
    \frac{1}{(d^2-1) \, (d^2 -2)}
  \end{pmatrix}
\end{equation}
This identity follows from 
\begin{gather*}
  \mbox{
    \raisebox{-.4cm}{
      \includegraphics[scale=0.6]{fmat2.21.eps}
    }
  }
  =
  \mbox{
    \raisebox{-.4cm}{
      \includegraphics[scale=0.6]{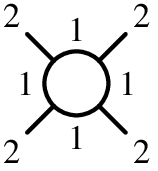}
    }
  }
  -\frac{1}{d} \,
  \mbox{
    \raisebox{-.3cm}{
      \includegraphics[scale=0.6]{skein3.21.eps}
    }
  }
  \\[2mm]
  \mbox{
    \raisebox{-.4cm}{
      \includegraphics[scale=0.6]{fmat2.22.eps}
    }
  }
  =
  \mbox{
    \raisebox{-.2cm}{
      \includegraphics[scale=0.6]{skein2.21.eps}
    }
  }
  -
  \frac{d^2}{\Delta_3} \,
  \mbox{
    \raisebox{-.4cm}{
      \includegraphics[scale=0.6]{fmat2.60.eps}
    }
  }
  +
  \frac{d}{\Delta_2 \, \Delta_3} \,
  \mbox{
    \raisebox{-.3cm}{
      \includegraphics[scale=0.6]{skein3.21.eps}
    }
  }
\end{gather*}
which can be derived after some algebra.

Actions of the braid operators on
bases
$\{ |0\rangle, |1\rangle, |2 \rangle \}$~\eqref{1-qubit_spin_1}
are computed similarly.
Representations of $\sigma_1$ and $\sigma_3$ can be given
from~\eqref{twist_formula} straightforwardly,
and that of $\sigma_2$ can be given once we use the
$F$-matrix~\eqref{F_matrix_22_22}.
Explicitly they are written as
\begin{equation}
  \label{braid_qubit_spin-1}
  \begin{gathered}
    \rho(\sigma_1) =
    \rho(\sigma_3) =
%    R_{12}= R_{34} =
    \diag \left(
      A^{-8} , - A^{-4} , A^4
    \right)
    \\[2mm]
    \rho(\sigma_2) 
%    R_{23}
    =
    \begin{pmatrix}
      \frac{A^8}{d^2-1} &
      - \frac{A^4}{\sqrt{d^2 - 1}} &
      \frac{\sqrt{d^4 - 3 \, d^2 +1}}{A^4 \, \left( d^2-1 \right)}
      \\[2mm]
      - \frac{A^4}{\sqrt{d^2 - 1}} &
      1- \frac{1}{d^2 -2} &
      \frac{1}{A^8 \, \left(d^2 -2 \right)} \,
      \sqrt{
        \frac{
          d^4 - 3 \, d^2+1}{d^2-1}
      }
      \\[2mm]
      \frac{\sqrt{d^4 - 3 \, d^2 +1}}{A^4 \, \left( d^2-1 \right)} &
      \frac{1}{A^8 \, \left(d^2 -2 \right)} \,
      \sqrt{
        \frac{
          d^4 - 3 \, d^2+1}{d^2-1}
      }
%       \frac{\sqrt{d^4 - 3 \, d^2 +1}}{A^8 \, \sqrt{d^2 - 1} \,
%         (d^2-2)}
      &
      \frac{1}{A^{16} \, \left( d^2-1 \right) \, \left( d^2-2 \right)}
    \end{pmatrix}
  \end{gathered}
\end{equation}
See that the braid relation~\eqref{Artin_braid} is fulfilled, and that
\begin{equation*}
  \rho\left(
    \sigma_1 \,    \sigma_2 \,    \sigma_3 \,
    \sigma_3 \,    \sigma_2 \,    \sigma_1 \,
  \right)
  =
  A^{-16} \cdot \mathbf{1}
\end{equation*}
The twist~\eqref{theta_twist} is follows from~\eqref{twist_formula}
because
all the quasi-particles are spin-$1$;
\begin{equation}
  \label{twist_spin-1}
  \rho(\theta)
  = A^8
\end{equation}

%%%%%
\subsection{5-Quasi-Particles States}

In $SU(2)_K$ theory with $K\geq 5$, the dimension of the Hilbert space
of 5-quasi particles is 6.
We choose  bases as
\begin{equation}
  \label{state_5_quasi}
  \begin{aligned}
    |01 \rangle
    & =
    \frac{d}{\Delta_2 \, \sqrt{\Delta_3}} \,
    \mbox{
      \raisebox{-.3cm}{
        \includegraphics[scale=0.7]{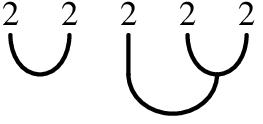}
      }
    }
    \\
    |10 \rangle
    & =
    \frac{d}{\Delta_2 \, \sqrt{\Delta_3}} \,
    \mbox{
      \raisebox{-.3cm}{
        \includegraphics[scale=0.7]{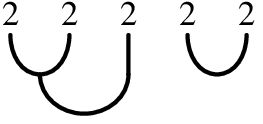}
      }
    }
    \\
    |11 \rangle
    & =
    \frac{d^3}{
      \Delta_3 \, \sqrt{ \Delta_2 \,    \Delta_3 
      }} \,
    \mbox{
      \raisebox{-4mm}{
        \includegraphics[scale=0.7]{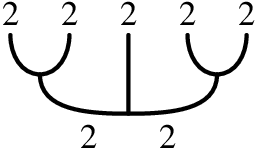}
      }
    }
    \\
    |12 \rangle
    & =
    \frac{d}{
      \sqrt{\Delta_3 \, \Delta_4}
    } \,
    \mbox{
      \raisebox{-4mm}{
        \includegraphics[scale=0.7]{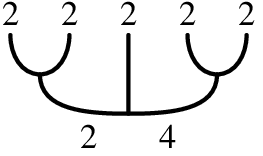}
      }
    }
    \\
    |21 \rangle
    & =
    \frac{d}{
      \sqrt{\Delta_3 \, \Delta_4}
    } \,
    \mbox{
      \raisebox{-4mm}{
        \includegraphics[scale=0.7]{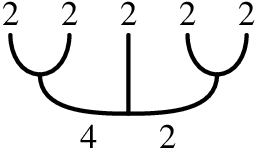}
      }
    }
    \\
    |22 \rangle
    & =
    \sqrt{\frac{
        d\, \Delta_3
      }{\Delta_4 \, \Delta_5}
    } \,
    \mbox{
      \raisebox{-4mm}{
        \includegraphics[scale=0.7]{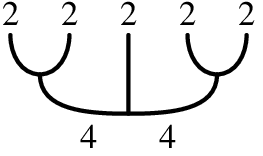}
      }
    }
  \end{aligned}
\end{equation}
Here $|ab\rangle$ means that  intermediate states have spins-$ a$,
$b$ as can be seen from~\eqref{state_5_quasi},
and we give generalization 
later by use of the quantum $6j$-symbols.
We have normalized these states to be
$\left\langle i | j \right\rangle =\delta_{i,j}$, where dual
states $\left\langle i \right|$ are
defined using upside down trivalent graph in~\eqref{state_5_quasi} 
with the same
normalization factors.

To obtain the representation of the braid operators,
we need another $F$-matrix beside~\eqref{F_matrix_22_22};
\begin{equation}
  \begin{pmatrix}
    \mbox{
      \raisebox{-.2cm}{
        \includegraphics[scale=0.6]{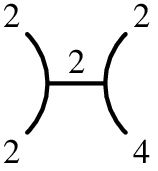}
      }
    }
    \\[2mm]
    \mbox{
      \raisebox{-.2cm}{
        \includegraphics[scale=0.6]{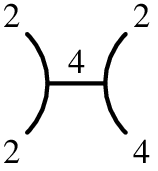}
      }
    }
  \end{pmatrix}
  =
  \mathbf{F}_{24}^{22} \,
  \begin{pmatrix}
    \mbox{
      \raisebox{-.2cm}{
        \includegraphics[scale=0.6]{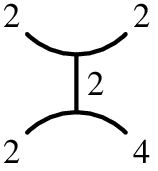}
      }
    }
    \\[2mm]
    \mbox{
      \raisebox{-.2cm}{
        \includegraphics[scale=0.6]{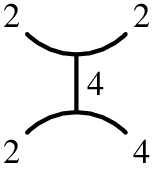}
      }
    }
  \end{pmatrix}
\end{equation}
where
\begin{equation}
  \mathbf{F}_{24}^{22}
  =
  \begin{pmatrix}
    -\frac{d}{\Delta_3} & 1 \\[2mm]
    1 - \frac{d^2}{\left( \Delta_3 \right)^2} & \frac{d}{\Delta_3}
  \end{pmatrix}
\end{equation}

A twist formula~\eqref{twist_formula} gives
representations on spaces spanned by
$\left\{
  \left| 01 \right\rangle,
  \left| 10 \right\rangle,
  \left| 11 \right\rangle,
  \left| 12 \right\rangle,
  \left| 21 \right\rangle,
  \left| 22 \right\rangle
\right\}$;
\begin{equation}
  \label{braid_5_spin-1-1}
  \begin{gathered}
    \rho(\sigma_1)
    =
    \diag
    \left(
      A^{-8} , - A^{-4} , - A^{-4} , - A^{-4} , A^4, A^4
    \right)
    \\[2mm]
    \rho(\sigma_4)
    =
    \diag
    \left(
      - A^{-4} , A^{-8} , - A^{-4} , A^4 , - A^{-4} , A^4
    \right)
  \end{gathered}
\end{equation}
Using above mentioned $F$-matrices, we obtain
\begin{equation}
  \label{braid_5_spin-1-2}
  \begin{gathered}
    \rho(\sigma_2)
    =
    \begin{pmatrix}
      \frac{A^8}{\Delta_2} & 0 & - \frac{A^4}{\sqrt{\Delta_2}} & 0
      & \frac{\sqrt{\Delta_4}}{A^4 \, \Delta_2} & 0
      \\[2mm]
      0 & - A^{-4} & 0 & 0 & 0 & 0
      \\[2mm]
      - \frac{A^4}{\sqrt{\Delta_2}} & 0 &
      \frac{\Delta_5}{\Delta_2  \,\Delta_3} & 0
      &  \frac{d}{A^8 \, \Delta_3} \,
      \sqrt{\frac{\Delta_4}{\Delta_2}}
      & 0
      \\[2mm]
      0 & 0 & 0 & A^8 \, \frac{d}{\Delta_3} & 0
      & \frac{\sqrt{d \, \Delta_5}}{\Delta_3}
      \\[2mm]
      \frac{\sqrt{\Delta_4}}{A^4 \, \Delta_2} & 0
      &   \frac{d}{A^8 \, \Delta_3} \,
      \sqrt{\frac{\Delta_4}{\Delta_2}} & 0
      & \frac{d}{A^{16} \, \Delta_2 \, \Delta_3} & 0
      \\[2mm]
      0 & 0 & 0 & \frac{\sqrt{d \, \Delta_5}}{\Delta_3} & 0
      &
      - \frac{d}{A^8 \, \Delta_3}
    \end{pmatrix}
    \\[2mm]
    \rho(\sigma_3)
    =
    \begin{pmatrix}
      - A^{-4} & 0 & 0 & 0 & 0 & 0 
      \\[2mm]
      0 &
      \frac{A^8}{\Delta_2} & - \frac{A^4}{\sqrt{\Delta_2}} 
      & \frac{\sqrt{\Delta_4}}{A^4 \, \Delta_2} & 0& 0
      \\[2mm]
      0 &
      - \frac{A^4}{\sqrt{\Delta_2}} & 
      \frac{\Delta_5}{\Delta_2  \,\Delta_3} 
      &  \frac{d}{A^8 \, \Delta_3} \,
      \sqrt{\frac{\Delta_4}{\Delta_2}} & 0
      & 0
      \\[2mm]
      0 &\frac{\sqrt{\Delta_4}}{A^4 \, \Delta_2} 
      &   \frac{d}{A^8 \, \Delta_3} \,
      \sqrt{\frac{\Delta_4}{\Delta_2}} 
      & \frac{d}{A^{16} \, \Delta_2 \, \Delta_3} & 0& 0
      \\[2mm]
      0& 0 & 0 & 0 & A^8 \, \frac{d}{\Delta_3} 
      & \frac{\sqrt{d \, \Delta_5}}{\Delta_3}
      \\[2mm]
      0& 0 & 0 & 0 & \frac{\sqrt{d \, \Delta_5}}{\Delta_3} 
      &
      - \frac{d}{A^8 \, \Delta_3}
    \end{pmatrix}
  \end{gathered}
\end{equation}

We can see that these satisfies~\eqref{Artin_braid}, and
\begin{equation*}
  \rho \left(
    \sigma_1 \,    \sigma_2 \,    \sigma_3 \,    \sigma_4 \,
    \sigma_4 \,    \sigma_3 \,    \sigma_2 \,    \sigma_1 \,
  \right) = A^{-16} \cdot \mathbf{1}
\end{equation*}
The twist is represented as~\eqref{twist_spin-1}.

%%%%%
\subsection{Topological Entanglement Entropy of Spin-1 Quasi-Particle States}

% The topological entanglement entropy~\eqref{von_Neumann_entropy} can
% be computed in the same manner.
% We study the topological entanglement
% entropy~\eqref{topological_entanglement} for quasi-particle states
% with spin-$1$.

\subsubsection{2-Quasi-Particle State}
We start from  the 2-quasi-particle state
$\left|\text{qp}_2\right\rangle$~\eqref{qp_2_spin-1}.
Here
we suppose that the first quasi-particle belongs to Alice and the
second to Bob;
$A=\{1 \}$ and $B=\{ 2 \}$.
By use of~\eqref{depict_rho_A}
the reduced density matrix  is depicted as
\begin{equation}
  \rho_{\{ 1\}} =
  \frac{1}{d^2 -1 } \,
  \mbox{
    \raisebox{-.8cm}{
      \includegraphics[scale=0.8]{proj1.14.eps}
    }
  }
%   \mbox{
%     \raisebox{-.3cm}{
%       \includegraphics[scale=0.8]{qu5.11.eps}
%     }
%   }
\end{equation}
from which~\eqref{depict_S_A} gives
\begin{equation}
  S_{\{ 1 \}}^{\text{topo}}
  =
  S_{\{ 1 \}}
  =
  \log \left( d^2-1 \right)
\end{equation}
Recalling the topological entanglement entropy~\eqref{qp_S_1_spin-12}
of the  two spin-$1/2$ quasi-particle states,
we see that $d$ is replaced with $d^2 -1$ which is nothing but
the quantum
dimension of the spin-$1$-quasi-particle who intertwines Alice and Bob.

\subsubsection{3-Quasi-Particle State}
In the 3-quasi-particle state
$\left|\text{qp}_3\right\rangle$~\eqref{qp_3_spin-1}, we 
suppose that
the owner of the
first and the second quasi-particles is Alice
while the third is Bob;
$A=\{1,2 \}$ and
$B=\{3 \}$.
Taking a trace, \emph{i.e.}, connecting both ends of the Wilson line
of the third particle~\eqref{depict_rho_A}, we get
\begin{equation}
  \rho_{\{1 ,2 \} }
  =
  \frac{d}{
    \left(d^2 -1  \right) \,
    \left(d^2 -2  \right)
  } \,
  \mbox{
    \raisebox{-8mm}{
      \includegraphics[scale=0.8]{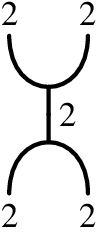}
    }
  }
\end{equation}
By definition~\eqref{depict_S_A},
we get the topological entanglement entropy
\begin{equation}
  S_{\{1 ,2 \} }^{\text{topo}}
  =
  S_{\{1 ,2 \} } =
  \log \left( d^2 -1 \right)
\end{equation}
This agrees with the logarithm of the quantum dimension of the
spin-$1$ quasi-particle which intertwines Alice's and Bob's quasi-particles.

%%%%%
\subsubsection{4-Quasi-Particle State}

We compute the entanglement entropy for the 4-quasi-particle states
defined by
\begin{equation}
  \label{state_Psi_spin-1}
  \left| \Psi \right\rangle
  =
  p_0 \, \left| 0 \right\rangle
  +
  p_1 \, \left| 1 \right\rangle
  +
  p_2 \, \left| 2 \right\rangle
\end{equation}
Here
bases
$ \left| a\right\rangle$ are
defined in~\eqref{1-qubit_spin_1}, and
$
\left|p_0 \right|^2+
\left|p_1 \right|^2+
\left|p_2 \right|^2
=1$.
As the first example,
we group four quasi-particles as
$A=\{1,2 \}$ and
$B=\{3,4\}$.
Through~\eqref{depict_rho_A} we get the Alice's reduced density matrix
as
\begin{equation}
  \rho_{\{1, 2 \} }
  =
  \frac{
    \left|p_0\right|^2}{
    d^2 -1
  } \,
  \mbox{
    \raisebox{-6mm}{
      \includegraphics[scale=0.8]{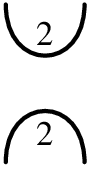}
    }
  }
  +
  \frac{
    \left|p_1\right|^2 \, d}{
    \left(d^2 -1 \right) \,     \left(d^2 -2 \right) 
  } \,
  \mbox{
    \raisebox{-8mm}{
      \includegraphics[scale=0.8]{reduced.6.eps}
    }
  }
  +
  \frac{
    \left|p_2\right|^2}{
    d^4 - 3 \, d^2 + 1
  } \,
  \mbox{
    \raisebox{-8mm}{
      \includegraphics[scale=0.8]{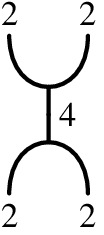}
    }
  }
\end{equation}
The entanglement entropy~\eqref{depict_S_A} can be computed
as
\begin{equation}
  S_{\{1,2 \}}
  =
  -
  \left| p_0 \right|^2
  \log  \left| p_0 \right|^2
  -
  \left| p_1 \right|^2
  \log 
  \left(
    \frac{\left| p_1 \right|^2}{
      d^2-1
    }
  \right)
  -
  \left| p_2 \right|^2
  \log  \left(
    \frac{\left| p_2 \right|^2}{
      d^4 - 3 \, d^2 +1
    }
  \right)
\end{equation}
As bases~\eqref{1-qubit_spin_1} have a form of~\eqref{orthogonal_base} due
to~\eqref{trivalent_theta}, the topological entanglement
entropy~\eqref{topological_entanglement} is given by
\begin{equation}
  S_{\{1,2\}}^{\text{topo}}
  =
  \left| p_1 \right|^2
  \log 
  \left(
    d^2 -1
  \right)
  +
  \left| p_2 \right|^2
  \log  \left(
    d^4 - 3 \, d^2 +1
  \right)
\end{equation}
We see the appearance of  quantum dimension
$d_2 = d^2 -1 $ and
$d_4 = d^4 -3 \, d^2 +1$ of quasi-particles
which respectively intertwines Alice and
Bob in the state 
$|1\rangle$ and $|2\rangle$.

When we assume that
$A=\{2,3\}$ and $B=\{1,4\}$, it is convenient to change bases using
the $F$-matrix~\eqref{F_matrix_22_22}.
We have another expression of~\eqref{state_Psi_spin-1};
\begin{multline}
  \label{another_Psi_spin-1}
  \left| \Psi \right\rangle
  =
  \widetilde{p_0} \, \frac{1}{d^2-1} \,
  \mbox{
    \raisebox{-.3cm}{
      \includegraphics[scale=0.8]{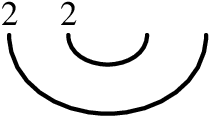}
    }
  }
  +
  \widetilde{p_1} \,
  \frac{d}{\left( d^2-2 \right) \, \sqrt{d^2-1}} \,
  \mbox{
    \raisebox{-.3cm}{
      \includegraphics[scale=0.8]{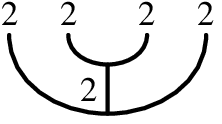}
    }
  }
  \\
  +
  \widetilde{p_2} \,
  \frac{1}{\sqrt{d^4 -3 \, d^2 +1}} \,
  \mbox{
    \raisebox{-.3cm}{
      \includegraphics[scale=0.8]{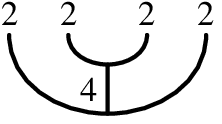}
    }
  }
\end{multline}
where
\begin{equation}
  \begin{aligned}
    \widetilde{p_0}
    & =
    p_0 \, \frac{1}{d^2-1} + p_1 \, \frac{1}{\sqrt{d^2-1}}
    +
    p_2 \, \frac{
      \sqrt{d^4 -3 \, d^2 +1}
    }{d^2-1}
    \\[2mm]
    \widetilde{p_1}
    & =
    p_0 \, \frac{1}{\sqrt{d^2-1}}
    + p_1 \, \frac{d^2-3}{d^2-2}
    -
    p_2 \,
    \frac{1}{d^2-2} \,
    \sqrt{
      \frac{d^4 -3 \, d^2 +1}{d^2-1}
    }
    \\[2mm]
    \widetilde{p_2}
    & =
    p_0 \,
    \frac{\sqrt{d^4-3 \, d^2 + 1}}{d^2-1} 
    - p_1 \, \frac{1}{d^2-2} \,
    \sqrt{
      \frac{d^4 -3 \, d^2 +1}{d^2 -1}
    }
    +
    p_2 \, \frac{
      1
    }{\left(d^2-1 \right) \, \left( d^2-2 \right)}
  \end{aligned}
\end{equation}
See that
$
\left| \widetilde{p_0} \right|^2 +
\left| \widetilde{p_1} \right|^2 +
\left| \widetilde{p_2} \right|^2 
=1$.
By the same computation,
we obtain the (topological) entanglement entropy as follows;
\begin{gather}
  S_{\{2,3 \}}
  =
  -
  \left| \widetilde{p_0} \right|^2
  \log  \left| \widetilde{p_0} \right|^2
  -
  \left| \widetilde{p_1} \right|^2
  \log 
  \left(
    \frac{\left| \widetilde{p_1} \right|^2}{
      d^2-1
    }
  \right)
  -
  \left| \widetilde{p_2} \right|^2
  \log  \left(
    \frac{\left| \widetilde{p_2} \right|^2}{
      d^4 - 3 \, d^2 +1
    }
  \right)
  \\[2mm]
  S_{\{2,3\}}^{\text{topo}}
  =
  \left| \widetilde{p_1} \right|^2
  \log 
  \left(
    d^2 -1
  \right)
  +
  \left| \widetilde{p_2} \right|^2
  \log  \left(
    d^4 - 3 \, d^2 +1
  \right)
\end{gather}
which is consistent with our interpretation of the topological
entanglement entropy;
when we look at~\eqref{another_Psi_spin-1},
we find that
parameters $\widetilde{p_1}$ and $\widetilde{p_2}$
respectively denote probabilities that
the spin-$1$ and spin-$2$ quasi-particle intertwines Alice and Bob.

%%%%%
\subsection{Fibonacci Anyon}

The spin-1 quasi-particle in the $SU(2)_3$ theory
is known as the Fibonacci anyon~\cite{JPresk04lecture}.
This is because the dimension of the Hilbert space of
$n$-quasi-particle with spin-$1$, \emph{i.e.}, the number of paths from
$(0,0)$ to $(n,0)$ in Fig.~\ref{fig:Bratteli_spin-1}, coincides with
the Fibonacci number.
For instance,
among three states of 4-quasi-particles~\eqref{1-qubit_spin_1},
the state $|2 \rangle$ is
not admissible~\eqref{admissible};
the norm of the diagram for $|2\rangle$ vanishes,
and it is a null state.
Setting $A=\I \, \E^{\pi \I/10}$ and
$d$ to be the golden-ratio
$d=2 \cos(\pi/5)=\frac{1+\sqrt{5}}{2} \equiv \tau$
in~\eqref{braid_qubit_spin-1}
we obtain 2-dimensional representation  of the braid operators on
$\left\{ |0\rangle, |1\rangle \right\}$ as
\begin{equation}
  \label{braiding_Fibonacci_1}
  \begin{gathered}
    \rho(\sigma_1) =
    \rho(\sigma_3) 
%    R_{12} = R_{34}
    =
    \begin{pmatrix}
      \E^{-\frac{4}{5} \pi \I} & 0 \\[2mm]
      0 & \E^{\frac{3}{5} \pi \I}
    \end{pmatrix}
    \\[2mm]
    \rho(\sigma_2) =
%    R_{23} =
    \begin{pmatrix}
      \frac{\E^{\frac{4}{5} \pi \I}}{\tau} &
      -  \frac{\E^{\frac{2}{5} \pi \I}}{\sqrt{\tau}}
      \\[2mm]
      -  \frac{\E^{\frac{2}{5} \pi \I}}{\sqrt{\tau}} &
      -\frac{1}{\tau}
    \end{pmatrix}
    \\[2mm]
    \rho(\theta)=
    \E^{\frac{4}{5} \pi \I}
  \end{gathered}
\end{equation}
which are given in Ref.~\citenum{JPresk04lecture}
assuming the unitarity condition.
% where $\tau$ is the golden-ratio,
% $\tau=\frac{1+\sqrt{5}}{2}$~\cite{JPresk04lecture}.

A dimension of the Hilbert space of 5-quasi-particle states becomes 3;
the norms of diagrams in
the states
$\left|1 2 \right\rangle$,
$\left| 2 1 \right\rangle$, and
$\left| 2 2 \right\rangle$
vanish due to $\Delta_4=0$, and these states are not
admissible~\eqref{admissible}.
Note that only $\left| 2 2 \right\rangle$ is forbidden for a case of
$K=4$.
Then on bases $\{ |01\rangle, |1 0 \rangle , |1 1\rangle \}$, the
braiding matrices~\eqref{braid_5_spin-1-1}
and~\eqref{braid_5_spin-1-2}
reduce to
\begin{equation}
  \label{braiding_Fibonacci_2}
  \begin{gathered}
    \rho(\sigma_1)
    =
    \diag
    \left(
      \E^{-\frac{4}{5} \pi \I} ,
      \E^{\frac{3}{5} \pi \I} ,
      \E^{\frac{3}{5} \pi \I} 
    \right)
    \\[2mm]
    \rho(\sigma_2)
    =
    \begin{pmatrix}
      \frac{\E^{\frac{4}{5} \pi \I}}{\tau} & 0 &
      - \frac{\E^{\frac{2}{5} \pi \I}}{\sqrt{\tau}}
      \\
      0 & \E^{\frac{3}{5} \pi \I} & 0
      \\
      -\frac{\E^{\frac{2}{5} \pi \I}}{\sqrt{\tau}} & 0 &
      -\frac{1}{\tau}
    \end{pmatrix}
    \\[2mm]
    \rho(\sigma_3)
    =
    \begin{pmatrix}
      \E^{\frac{3}{5} \pi \I}  & 0 & 0 
      \\
      0 & \frac{\E^{\frac{4}{5} \pi \I}}{\tau} & 
      - \frac{\E^{\frac{2}{5} \pi \I}}{\sqrt{\tau}}
      \\
      0 & -\frac{\E^{\frac{2}{5} \pi \I}}{\sqrt{\tau}} & 
      -\frac{1}{\tau}
    \end{pmatrix}
    \\[2mm]
    \rho(\sigma_4)
    =
    \diag
    \left(
      \E^{\frac{3}{5} \pi \I} ,
      \E^{-\frac{4}{5} \pi \I} ,
      \E^{\frac{3}{5} \pi \I} 
    \right)
  \end{gathered}
\end{equation}

On the  topological entanglement
entropy~\eqref{topological_entanglement}, we recall that the quantum
dimension of quasi-particles with spin-$1/2$, $1$, and $3/2$ are
respectively given by
$d_1=\tau$, $d_2=\tau$, and $d_3=1$.
Thus the topological entanglement entropy always takes a form
of $\left|p_\tau\right|^2 \log \tau$
where
$\left| p_\tau \right|^2$,
satisfying
$0 \leq \left|p_\tau\right|^2 \leq 1$,
is a probability that Alice and Bob are
intertwined through quasi-particle with spin-$1/2$ or spin-$1$.

%%%%
%%%% 
\section{Qubit from Different Species of Quasi-particles}
\label{sec:different}

In previous sections, we have only considered states which are
constructed from quasi-particles with same spin;
spin-$1/2$ and spin-$1$.
In this section we shall propose another candidate of topological
qubit which is composed from
4-quasi-particles  with different spins.
Namely we set $1 < n \leq K-1$, and
study states constituted from two spin-$1/2$ quasi-particles and two
spin-$n/2$ quasi-particles.
We set bases as follows;
\begin{equation}
  \label{2d_base_1_n}
  \begin{aligned}
    |0_1 \rangle
    &=
    \frac{
      \sqrt{\Delta_{n-1}}
    }{\Delta_n} \,
    \mbox{
      \raisebox{-.4cm}{
        \includegraphics[scale=0.7]{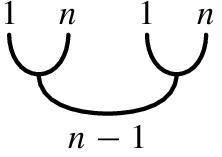}
      }
    }
    & \hspace{16mm}
    |1_1 \rangle 
    & =
    \frac{1}{
      \sqrt{\Delta_{n+1}}
    } \,
    \mbox{
      \raisebox{-.4cm}{
        \includegraphics[scale=0.7]{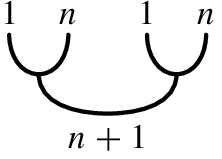}
      }
    }
    \\[2mm]
    |0_2 \rangle
    &=
    \frac{
      \sqrt{\Delta_{n-1}}
    }{\Delta_n} \,
    \mbox{
      \raisebox{-.4cm}{
        \includegraphics[scale=0.7]{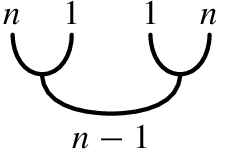}
      }
    }
    &
    |1_2 \rangle 
    & =
    \frac{1}{
      \sqrt{\Delta_{n+1}}
    } \,
    \mbox{
      \raisebox{-.4cm}{
        \includegraphics[scale=0.7]{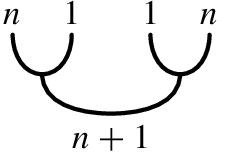}
      }
    }
    \\[2mm]
    |0_3 \rangle
    &=
    \frac{
      \sqrt{\Delta_{n-1}}
    }{\Delta_n} \,
    \mbox{
      \raisebox{-.4cm}{
        \includegraphics[scale=0.7]{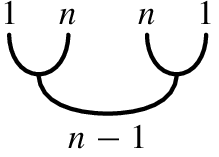}
      }
    }
    &
    |1_3 \rangle 
    & =
    \frac{1}{
      \sqrt{\Delta_{n+1}}
    } \,
    \mbox{
      \raisebox{-.4cm}{
        \includegraphics[scale=0.7]{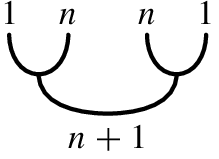}
      }
    }
    \\[2mm]
    |0_4 \rangle
    &=
    \frac{
      \sqrt{\Delta_{n-1}}
    }{\Delta_n} \,
    \mbox{
      \raisebox{-.4cm}{
        \includegraphics[scale=0.7]{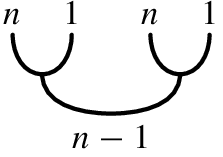}
      }
    }
    &
    |1_4 \rangle 
    & =
    \frac{1}{
      \sqrt{\Delta_{n+1}}
    } \,
    \mbox{
      \raisebox{-.4cm}{
        \includegraphics[scale=0.7]{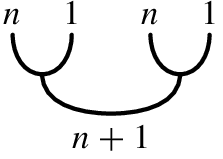}
      }
    }
    \\[2mm]
    |0_5 \rangle 
    & =
    \frac{1}{\sqrt{d \, \Delta_n}} \,
    \mbox{
      \raisebox{-.2cm}{
        \includegraphics[scale=0.7]{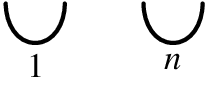}
      }
    }
    &
    |1_5 \rangle
    & =
    \sqrt{
      \frac{d \, \Delta_{n-1}}{\Delta_n \, \Delta_{n+1}}
    } \,
    \mbox{
      \raisebox{-.4cm}{
        \includegraphics[scale=0.7]{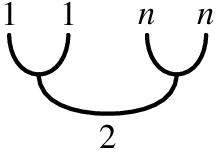}
      }
    }
    \\[2mm]
    |0_6 \rangle 
    & =
    \frac{1}{\sqrt{d \, \Delta_n}} \,
    \mbox{
      \raisebox{-.2cm}{
        \includegraphics[scale=0.7]{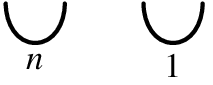}
      }
    }
    &
    |1_6 \rangle
    & =
    \sqrt{
      \frac{d \, \Delta_{n-1}}{\Delta_n \, \Delta_{n+1}}
    } \,
    \mbox{
      \raisebox{-.4cm}{
        \includegraphics[scale=0.7]{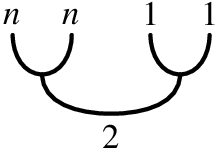}
      }
    }
  \end{aligned}
\end{equation}
Dual states $\left\langle i_a \right|$ are upside down of
$\left| i_a \right\rangle$ with the same normalization  factor.
A condition~\eqref{Delta_special}
proves that
these states are physical states,
\emph{i.e.},
norm of each diagram  is positive.
Normalizations are chosen so that
\begin{equation*}
  \left\langle i_a \middle| j_a \right\rangle=\delta_{i,j}
\end{equation*}
for $i,j \in \{0,1\}$ and $1 \leq a \leq 6$.

To compute explicit representation of braid operators, we prepare the
$F$-matrices.
Due to the definition of the idempotent, we find
\begin{gather}
  \begin{pmatrix}
        \mbox{
      \raisebox{-.2cm}{
        \includegraphics[scale=0.6]{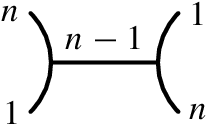}
      }
    }
    \\[2mm]
    \mbox{
      \raisebox{-.2cm}{
        \includegraphics[scale=0.6]{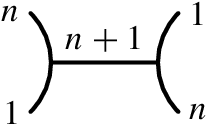}
      }
    }
  \end{pmatrix}
  =
  \begin{pmatrix}
    \frac{(-1)^{n+1}}{\Delta_n} & 1 
    \\[2mm]
    1 - \frac{1}{\left(\Delta_n \right)^2} &
    \frac{(-1)^n}{\Delta_n}
  \end{pmatrix}
  \,
  \begin{pmatrix}
    \mbox{
      \raisebox{-.2cm}{
        \includegraphics[scale=0.6]{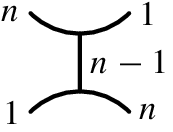}
      }
    }
    \\[2mm]
    \mbox{
      \raisebox{-.2cm}{
        \includegraphics[scale=0.6]{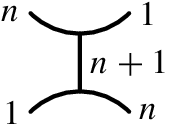}
      }
    }
  \end{pmatrix}
  \\[2mm]
  \begin{pmatrix}
    \mbox{
      \raisebox{-.2cm}{
        \includegraphics[scale=0.6]{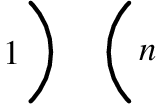}
      }
    }
    \\[2mm]
    \mbox{
      \raisebox{-.2cm}{
        \includegraphics[scale=0.6]{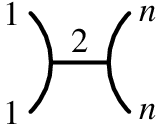}
      }
    }
  \end{pmatrix}
  =
  \begin{pmatrix}
    \frac{\Delta_{n-1}}{\Delta_n} & 1
    \\[2mm]
    \frac{\Delta_{n+1}}{d \, \Delta_n} &
    - \frac{1}{d}
  \end{pmatrix}
  \,
  \begin{pmatrix}
    \mbox{
      \raisebox{-.2cm}{
        \includegraphics[scale=0.6]{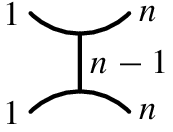}
      }
    }
    \\[2mm]
    \mbox{
      \raisebox{-.2cm}{
        \includegraphics[scale=0.6]{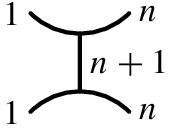}
      }
    }
  \end{pmatrix}
\end{gather}
Using these $F$-matrices, 
we can compute
the actions of braid operators on
$\left\{ |0_1 \rangle, |1_1 \rangle,
  |0_2 \rangle, |1_2 \rangle,
  \cdots
  |0_6 \rangle, |1_6 \rangle
\right\}$
are computed.
The actions of $\sigma_1$ and $\sigma_3$ follow
from~\eqref{twist_formula} as
\begin{equation}
  \begin{gathered}
    \rho(\sigma_1)
    =
    \begin{pmatrix}
      0 & \mathbf{R}_1 & 0 & 0 & 0 & 0
      \\
      \mathbf{R}_1 & 0 & 0 & 0 & 0 & 0
      \\
      0 & 0 & 0 & \mathbf{R}_1 & 0 & 0
      \\
      0 & 0 & \mathbf{R}_1 & 0 & 0 & 0
      \\
      0 & 0 & 0 & 0 & \mathbf{R}_2 & 0
      \\
      0 & 0 & 0 & 0 & 0 & \mathbf{R}_3
    \end{pmatrix}
    \\[2mm]
    \rho(\sigma_3)
    =
    \begin{pmatrix}
      0 & 0 & \mathbf{R}_1 & 0 & 0 & 0
      \\
      0 & 0 & 0 & \mathbf{R}_1 & 0 & 0
      \\
      \mathbf{R}_1 & 0 & 0 & 0 & 0 & 0
      \\
      0 & \mathbf{R}_1 & 0 & 0 & 0 & 0
      \\
      0 & 0 & 0 & 0 & \mathbf{R}_3 & 0
      \\
      0 & 0 & 0 & 0 & 0 & \mathbf{R}_2
    \end{pmatrix}
  \end{gathered}
\end{equation}
where $0$ denotes a 2 by 2 null matrix, and
$R_a$ are 2 by 2 diagonal matrices defined by
\begin{equation*}
  \begin{gathered}
    \mathbf{R}_1
    =
    \diag \left( - A^{-n-2} , A^n \right)
    \\[2mm]
    \mathbf{R}_2
    =
    \diag \left(
      -A^{-3} , A
    \right)
    \\[2mm]
    \mathbf{R}_3
    =
    \diag \left(
      (-1)^n \, A^{-n^2 - 2 n},
      (-1)^{n+1} \, A^{-n^2 - 2 n +4}
    \right)
  \end{gathered}
\end{equation*}
Then by use of above $F$-matrices we obtain the representations of
$\sigma_2$ as
\begin{equation}
  \rho(\sigma_2 )
  =
  \begin{pmatrix}
    0 & 0 & 0 & 0 & \mathbf{B}_1 & 0
    \\
    0 & \mathbf{B}_2 & 0 & 0 & 0 & 0
    \\
    0 & 0 & \mathbf{B}_3 & 0 & 0 & 0
    \\
    0 & 0 & 0 & 0 & 0 & \mathbf{B}_1
    \\
    \mathbf{B}_1^{\mathrm{T}} & 0 & 0 & 0 & 0 & 0
    \\
    0 & 0 & 0 & \mathbf{B}_1^{\mathrm{T}} & 0 & 0
  \end{pmatrix}
\end{equation}
where $0$ is a 2 by 2 null matrix as before, and $\mathbf{B}_a$ are
\begin{equation*}
  \begin{gathered}
    \mathbf{B}_1
    =
    \frac{1}{\sqrt{d \, \Delta_n}} \,
    \begin{pmatrix}
      - A^{n+2} \, \sqrt{\Delta_{n-1}} &
      A^{n-2} \, \sqrt{\Delta_{n+1}}
      \\[2mm]
      A^{-n} \, \sqrt{\Delta_{n+1}}
      &
      A^{-n-4} \, \sqrt{\Delta_{n-1}}
    \end{pmatrix}
    \\[2mm]
    \mathbf{B}_2
    =
    \frac{1}{A \, \Delta_n} \,
    \begin{pmatrix}
      A^2 \, \Delta_n + \Delta_{n-1} &
      \sqrt{\Delta_{n+1} \, \Delta_{n-1}}
      \\[2mm]
      \sqrt{\Delta_{n+1} \, \Delta_{n-1}}
      &
      - \left(
        A^{-2} \, \Delta_n + \Delta_{n-1}
      \right)
    \end{pmatrix}
    \\[2mm]
    \mathbf{B}_3
    =
    (-1)^{n+1} \, A^{-n^2 -2 n +3} \, \mathbf{B}_2
  \end{gathered}
\end{equation*}
We mean that $\mathbf{B}_1^{\mathrm{T}}$ is a transpose of $\mathbf{B}_1$.

We note that
$\rho(\sigma_a^2)$ for all $a$ are block diagonal
\begin{equation}
  \label{block_matrix_2}
  \begin{gathered}
    \rho\left(
      \sigma_1^{~2}
    \right)
    =
    \diag\left(
      \mathbf{R}_1^{~2},
      \mathbf{R}_1^{~2},
      \mathbf{R}_1^{~2},
      \mathbf{R}_1^{~2},
      \mathbf{R}_2^{~2},
      \mathbf{R}_3^{~2}
    \right)
    \\[2mm]
    \rho\left(
      \sigma_2^{~2}
    \right)
    =
    \diag\left(
      \mathbf{B}_1 \, \mathbf{B}_1^{\mathrm{T}},
      \mathbf{B}_2^{~2},
      \mathbf{B}_3^{~2},
      \mathbf{B}_1 \, \mathbf{B}_1^{\mathrm{T}},
      \mathbf{B}_1^{\mathrm{T}} \,      \mathbf{B}_1 ,
      \mathbf{B}_1^{\mathrm{T}} \,      \mathbf{B}_1 
    \right)
    \\[2mm]
    \rho\left(
      \sigma_3^{~2}
    \right)
    =
    \diag\left(
      \mathbf{R}_1^{~2},
      \mathbf{R}_1^{~2},
      \mathbf{R}_1^{~2},
      \mathbf{R}_1^{~2},
      \mathbf{R}_3^{~2},
      \mathbf{R}_2^{~2}
    \right)
  \end{gathered}
\end{equation}
This fact shows that,
when we restrict braiding operators generated from
even powers of braid operators
$\left\{ \sigma_1^{~2},
 \sigma_2^{~2},
 \sigma_3^{~2}
\right\}$,
all the 2-dimensional
spaces $\{ | 0_a \rangle , |1_a \rangle \}$  are preserved.
Then these 2-dimensional space may be
treated as qubit, and block matrices in~\eqref{block_matrix_2} are
treated as unitary operators on these spaces.

It should be noted that
other sets of braid operators preserves $2$-dimensional space,
\emph{e.g.}
\begin{itemize}
\item A set of
  $\left\{ \sigma_1^{~2}, \sigma_2, \sigma_3^{~2} \right\}$ preserves
  $|i_2\rangle$ and
  $|i_3 \rangle$.

\item A set of
  $\left\{ \sigma_1, \sigma_2^{~2}, \sigma_3 \right\}$ preserves
  $|i_5\rangle$ and
  $|i_6 \rangle$.

\end{itemize}

It is remarked that some of the braiding matrices are given in
Ref.~\citenum{ArdonSchou07a} based on explicit form of the correlation
functions.

%%%

%%%%%%%%
%\appendix
\section{Many-Quasi-Particle States and Quantum
  $6j$ Symbol}
\label{sec:6j}

Up to now, we have  derived braiding matrices for systems of small
numbers of quasi-particles
%of non-Abelian quantum Hall system
based on the
skein theory.
We have derived all $F$-matrices in
an  elementary and combinatorial way
based on the Jones--Wenzl idempotent.
% This elementary and combinatorial method 
% enables us to
% calculate the $F$-matrices, which are fundamental tools in our
% analysis,
% based on the Jones--Wenzl idempotent.
Although,
explicit forms of the general $F$-matrices
are known as the quantum $6j$ system, and
we can construct the
braiding matrices for arbitrary numbers of quasi-particles.

\subsection{Quantum $6j$ Symbol}
We borrow  results on the quantum $6j$ symbol from
Ref.~\citenum{KaufLins94Book}
(see also Refs.~\citenum{Licko97Book,CartFlatSait94a,MasbVoge94a}).
We define $q$-integer and $q$-factorial by
\begin{gather*}
  [ n ]
  = \frac{A^{2n} - A^{-2n}}{A^2 - A^{-2}}
  =
  (-1)^{n-1} \, \Delta_{n-1}
  \\[2mm]
  [n]! = \prod_{i=1}^n [i]
\end{gather*}

The quantum $6j$-symbol comes from the following
Temperley--Lieb recoupling diagram;
\begin{equation}
  \label{6j_symbol}
  \mbox{
    \raisebox{-.6cm}{
      \includegraphics[scale=0.88]{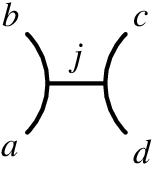}
    }
  }
  =
  \sum_i
  \begin{Bmatrix}
    a & b & i
    \\
    c & d & j
  \end{Bmatrix}
  \mbox{
    \raisebox{-.6cm}{
      \includegraphics[scale=0.88]{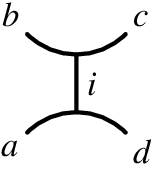}
    }
  }
\end{equation}
We have a complicated formula for the quantum $6j$-symbol,
which was first computed in Ref.~\citenum{KirilReshe89};
\begin{equation}
  \begin{Bmatrix}
    a & b & i
    \\
    c & d & j
  \end{Bmatrix}
  =
  \frac{\Delta_i}{\theta(a,d,i) \, \theta(b,c,i)} \,
  \Tet
  \begin{bmatrix}
    a & b & i
    \\
    c & d & j
  \end{bmatrix}
\end{equation}
Here
$\theta(a,b,c)$ is the $\theta$-net
\begin{equation}
  \theta(a,b,c)
  =
  \mbox{
    \raisebox{-8mm}{
      \includegraphics[scale=0.8]{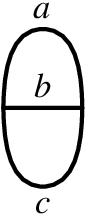}
    }
  }
\end{equation}
which is given explicitly as
\begin{equation}
  \theta(a,b,c)
  =
  (-1)^{i+j+k} \,
  \frac{
    [i+j+k+1]! \,
    [i]! \, [j]! \, [k]!
  }{
    [i+j]! \,
    [j+k]! \,
    [i+k]!
  }
\end{equation}
with
\begin{equation*}
  \begin{aligned}
    i
    & =
    \frac{-a+b+c}{2}
    & \hspace{8mm}
    j
    & =
    \frac{a-b+c}{2}
    & \hspace{8mm}
    k
    & =
    \frac{a+b-c}{2}
  \end{aligned}
\end{equation*}
The tetrahedral net is defined by
\begin{equation}
  \Tet
  \begin{bmatrix}
    A & B & E
    \\
    C & D & F
  \end{bmatrix}
  =
  \mbox{
    \raisebox{-9mm}{
      \includegraphics[scale=0.72]{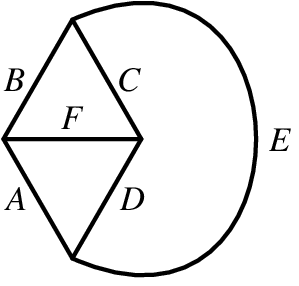}
    }
  }
\end{equation}
which is known to be
\begin{multline}
  \Tet
  \begin{bmatrix}
    A & B & E
    \\
    C & D & F
  \end{bmatrix}
  =
  \frac{\prod_{i,j}  [ b_j - a_i ]!}{
    [A ] ! \,    [B ] ! \,    [C ] ! \,
    [D ] ! \,    [E ] ! \,    [F ] ! 
  }
  \\
  \times
  \sum_{
    \max \{ a_i \} \leq s \leq \min \{b_j\}
  }
  \frac{
    (-1)^s \, [s+1]!
  }{
    \prod_i [s-a_i] ! \,
    \prod_j [b_j -s]!
  }
\end{multline}
with
\begin{equation*}
  \begin{aligned}
    a_1 & = \frac{1}{2}\,
    \left( A+ D+ E \right)
    \\
    a_2 & = \frac{1}{2}\,
    \left( B+ C+ E \right)
    \\
    a_3
    & = \frac{1}{2}\,
    \left( A+ B+ F \right)
    \\
    a_4 & = \frac{1}{2}\,
    \left( C+ D+ F \right)
  \end{aligned}
  \qquad \qquad
  \begin{aligned}
    b_1
    & = \frac{1}{2} \, \left(
      B+D+E+F \right)
    \\
    b_2
    & = \frac{1}{2} \, \left(
      A+C+E+F \right)
    \\
    b_3
    & = \frac{1}{2} \, \left(
      A+B+C+D \right)
  \end{aligned}
\end{equation*}

As a simple application  of the quantum $6j$-symbol~\eqref{6j_symbol},
we have
\begin{equation}
  \label{null_6j}
  \mbox{
    \raisebox{-.4cm}{
      \includegraphics[scale=0.88]{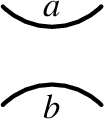}
    }
  }
  =
  \sum_{c}
  \frac{\Delta_c}{
    \theta(a,b,c)} \,
  \mbox{
    \raisebox{-.6cm}{
      \includegraphics[scale=0.88]{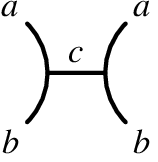}
    }
  }
\end{equation}
Note that the $\theta$-net appears from
\begin{equation}
  \label{trivalent_theta}
  \mbox{
    \raisebox{-7mm}{
      \includegraphics[scale=0.8]{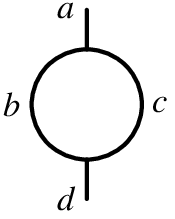}
    }
  }
  =
  \delta_{a,d} \, \frac{\theta(a,b,c)}{\Delta_a} \,
  \mbox{
    \raisebox{-.6cm}{
      \includegraphics[scale=0.8]{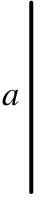}
    }
  }
\end{equation}

The quantum $6j$ symbol has symmetries;
\begin{gather}
  \label{symmetry_6j}
  \begin{Bmatrix}
    a & b & i
    \\
    c & d & j
  \end{Bmatrix}
  =
  \begin{Bmatrix}
    c & d & i
    \\
    a & b & j
  \end{Bmatrix}
\end{gather}
which follows from a $180^\circ$ rotation of~\eqref{6j_symbol}.
Also we have
\begin{equation}
  \label{symmetry_6j_2}
  \begin{aligned}
    \frac{\theta(a,d,i)}{\Delta_i} \,
    \begin{Bmatrix}
      a & b & i
      \\
      c & d & j
    \end{Bmatrix}
    & =
    \frac{\theta(a,b,j)}{\Delta_b} \,
    \begin{Bmatrix}
      j & c & b
      \\
      i & a & d
    \end{Bmatrix}
    \\
    & =
    \frac{\theta(c,d,j)}{\Delta_c} \,
    \begin{Bmatrix}
      d & i & c
      \\
      b & j & a
    \end{Bmatrix}
  \end{aligned}
\end{equation}
which comes from applications of the $6j$-symbol to the following net;
\begin{equation*}
  \mbox{
    \raisebox{-.6cm}{
      \includegraphics[scale=0.88]{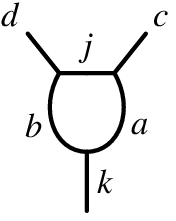}
    }
  }
\end{equation*}

\begin{figure}[tbhp]
  \centering
  \setlength{\unitlength}{8mm}
  \begin{picture}(16,9)
    \put(2,7){
      \mbox{
        \raisebox{-6mm}{
          \includegraphics[scale=0.8]{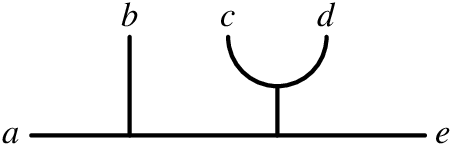}
        }
      }
    }    
    \put(7.5,7){\vector(1,0){1.2}}
    \put(9,7){
      \mbox{
        \raisebox{-6mm}{
          \includegraphics[scale=0.8]{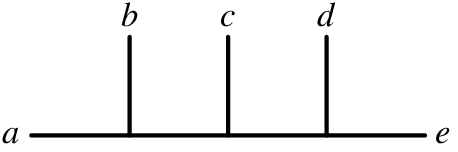}
        }
      }
    }    
    \put(4,6){\vector(-1,-2){0.5}}
    \put(0,4){
      \mbox{
        \raisebox{-6mm}{
          \includegraphics[scale=0.8]{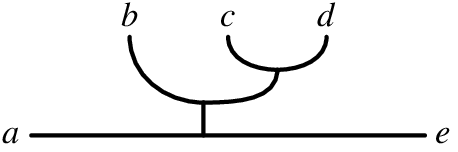}
        }
      }
    }    
    \put(12,6){\vector(1,-2){0.5}}
    \put(11,4){
      \mbox{
        \raisebox{-6mm}{
          \includegraphics[scale=0.8]{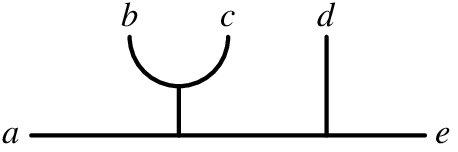}
        }
      }
    }    
    \put(3,3){\vector(1,-1){2}}
    \put(5,1){
      \mbox{
        \raisebox{-6mm}{
          \includegraphics[scale=0.8]{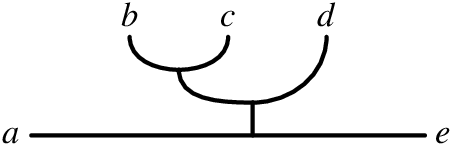}
        }
      }
    }    
    \put(11,1){\vector(1,1){2}}
  \end{picture}
  \caption{Pentagon identity}
  \label{fig:pentagon}
\end{figure}

Recursive use of the recoupling diagram~\eqref{6j_symbol} gives us the
orthogonality relation;
\begin{equation}
  \label{orthogonal_6j}
  \sum_{i}
  \begin{Bmatrix}
    a & b & i \\
    c & d & j
  \end{Bmatrix}
  \,
  \begin{Bmatrix}
    d & a & k \\
    b & c & i
  \end{Bmatrix}
  =
  \delta_{j,k}
\end{equation}
Well known is the pentagon equation, or the Biedenharn--Elliott
identity, which follows from
Fig.~\eqref{fig:pentagon};
\begin{equation}
  \sum_{m}
  \begin{Bmatrix}
    a & i & m \\
    d & e & j
  \end{Bmatrix}
  \,
  \begin{Bmatrix}
    b & c & \ell \\
    d & m & i
  \end{Bmatrix}
  \,
  \begin{Bmatrix}
    b & \ell & k \\
    e & a & m
  \end{Bmatrix}
  =
  \begin{Bmatrix}
    b & c & k \\
    j & a & i
  \end{Bmatrix}
  \,
  \begin{Bmatrix}
    k & c & \ell \\
    d & e & j
  \end{Bmatrix}
\end{equation}
Amongst others useful identity  we use~\cite{CartFlatSait94a}
\begin{gather}
  \label{6j_2_lambda}
  \left(
    \lambda_a^{d i}
  \right)^{-1} \,
  \lambda_j^{c d} \,
  \begin{Bmatrix}
    a & b & i 
    \\
    c & d & j
  \end{Bmatrix}
  =
  \sum_k
  \begin{Bmatrix}
    a & b & k
    \\
    d & c & j
  \end{Bmatrix} \,
  \left(
    \lambda_k^{b d}
  \right)^{-1} \,
  \begin{Bmatrix}
    d & b & i \\
    c & a & k
  \end{Bmatrix}
\end{gather}
which is the $B$-matrix~\cite{MoorSeib89a}.
To prove this, we  first  compute as
\begin{align*}
  \mbox{
    \raisebox{-.5cm}{
      \includegraphics[scale=0.88]{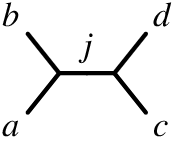}
    }
  }
  & =
  \lambda_j^{c d} \,
  \mbox{
    \raisebox{-.6cm}{
      \includegraphics[scale=0.88]{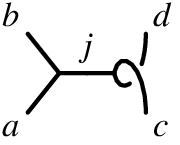}
    }
  }
  =
  \sum_i
  \lambda_j^{c d} \,
  \begin{Bmatrix}
    a & b & i
    \\
    c & d & j
  \end{Bmatrix}
  \,
  \mbox{
    \raisebox{-.6cm}{
      \includegraphics[scale=0.88]{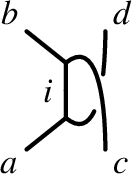}
    }
  }
\end{align*}
On the other hand, the left hand side is transformed as
\begin{align*}
  \mbox{
    \raisebox{-.5cm}{
      \includegraphics[scale=0.88]{q6ja.11.eps}
    }
  }
  & =
  \sum_k 
  \begin{Bmatrix}
    a & b & k
    \\
    d & c & j
  \end{Bmatrix}
  \,
  \mbox{
    \raisebox{-.6cm}{
      \includegraphics[scale=0.88]{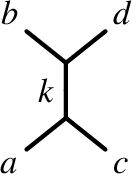}
    }
  }
  =
  \sum_k 
  \begin{Bmatrix}
    a & b & k
    \\
    d & c & j
  \end{Bmatrix}
  \,
  \left(
    \lambda_k^{b d}
  \right)^{-1}
  \,
  \mbox{
    \raisebox{-.6cm}{
      \includegraphics[scale=0.88]{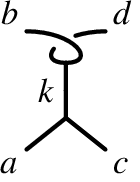}
    }
  }
  \\
  & =
  \sum_{i,k}
  \begin{Bmatrix}
    a & b & k
    \\
    d & c & j
  \end{Bmatrix}
  \,
  \left(
    \lambda_k^{b d}
  \right)^{-1}
  \,
  \begin{Bmatrix}
    c & a & i
    \\
    d & b & k
  \end{Bmatrix}
  \,
  \mbox{
    \raisebox{-.5cm}{
      \includegraphics[scale=0.88]{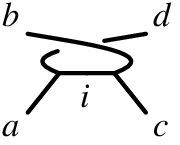}
    }
  }
  \\
  & =
  \sum_{i,k}
  \begin{Bmatrix}
    a & b & k
    \\
    d & c & j
  \end{Bmatrix}
  \,
  \left(
    \lambda_k^{b d}
  \right)^{-1}
  \,
  \begin{Bmatrix}
    d & b & i
    \\
    c & a & k
  \end{Bmatrix}
  \,
  \mbox{
    \raisebox{-.5cm}{
      \includegraphics[scale=0.88]{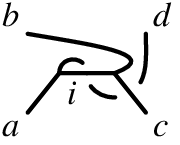}
    }
  }
  \\
  & =
  \sum_{i,k}
  \begin{Bmatrix}
    a & b & k
    \\
    d & c & j
  \end{Bmatrix}
  \,
  \left(
    \lambda_k^{b d}
  \right)^{-1}
  \,
  \begin{Bmatrix}
    d & b & i
    \\
    c & a & k
  \end{Bmatrix}
  \,
  \lambda_a^{d i} \,
  \mbox{
    \raisebox{-.5cm}{
      \includegraphics[scale=0.88]{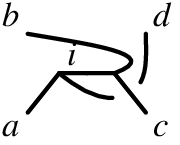}
    }
  }
\end{align*}
Combining these expressions, we obtain~\eqref{6j_2_lambda}.

%%%%%%%%% 
\subsection{Braid Relations for 4-Quasi-Particles with Arbitrary
  Spins}

We generalize  results in previous sections.
We define the state
\begin{equation}
  \label{4_qp_general}
  \left|
    i_{ab;cd}
  \right\rangle
  =
  \sqrt{
    \frac{ \Delta_i}{
      \theta(a,b,i) \,
      \theta(c,d,i)
    }
  } \,
  \mbox{
    \raisebox{-.5cm}{
      \includegraphics[scale=0.8]{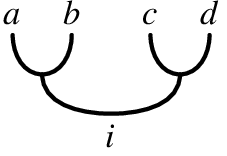}
    }
  }
\end{equation}
The dimension of the Hilbert space, \emph{i.e.}, the number of $i$ for fixed
$\{a, b, c, d\}$,
is the number of $i$'s such that
vertices $(a,b,i)$ and $(c,d,i)$ satisfy
the admissible condition~\eqref{admissible}.
The orthonormality follows from~\eqref{trivalent_theta};
\begin{equation*}
  \left\langle
    i_{ab;cd} 
    \middle|
    j_{ab;cd}
  \right\rangle = \delta_{i,j}
\end{equation*}

The braiding matrices on the  space spanned by
$\left|i_{ab;cd}\right\rangle$
can be computed by use of
the quantum $6j$-symbol~\eqref{6j_symbol} and
the twist formula~\eqref{twist_formula}.
For instance, we compute as follows;
\begin{align*}
  \sigma_2 
  \mbox{
    \raisebox{-.5cm}{
      \includegraphics[scale=0.7]{qubit.71.eps}
    }
  }
  & =
  \sum_k
  \begin{Bmatrix}
    a & b & k
    \\
    c & d & i
  \end{Bmatrix} \,
  \mbox{
    \raisebox{-0.6cm}{
      \includegraphics[scale=0.7]{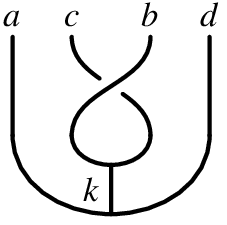}
    }
  }
  \\
  & =
  \sum_k
  \begin{Bmatrix}
    a & b & k
    \\
    c & d & i
  \end{Bmatrix} \,
  \left(
    \lambda_k^{b c}
  \right)^{-1}
  \,
  \mbox{
    \raisebox{-0.3cm}{
      \includegraphics[scale=0.7]{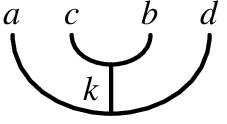}
    }
  }
  \\
  & =
  \sum_k
  \begin{Bmatrix}
    a & b & k
    \\
    c & d & i
  \end{Bmatrix} \,
  \left(
    \lambda_k^{b c}
  \right)^{-1}
  \,
  \begin{Bmatrix}
    c & b & j
    \\
    d & a & k
  \end{Bmatrix} \,
  \mbox{
    \raisebox{-0.5cm}{
      \includegraphics[scale=0.7]{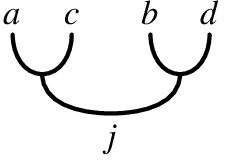}
    }
  }
\end{align*}
Applying~\eqref{6j_2_lambda} and using a normalization factor,
we obtain $\rho(\sigma_2)$.
In summary, we have the braiding matrices as
\begin{equation}
  \begin{gathered}
    \rho(\sigma_1) \,
    \left|i_{a b; c d} \right\rangle
    =
    \left( \lambda_i^{a b} \right)^{-1} \,
    \left|i_{b a; c d} \right\rangle
    \\[2mm]
    \rho(\sigma_2) \,
    \left|i_{a b; c d} \right\rangle
    =
    \sum_j
    \sqrt{
      \frac{
        \Delta_i \, \theta(a,c,j) \, \theta(b,d,j)
      }{
        \Delta_j \, \theta(a,b,i) \, \theta(c,d,i)
      }
    } \,
    \left(
      \lambda_a^{c j}
    \right)^{-1} \,
    \lambda_i^{c d} \,
    \begin{Bmatrix}
      a & b & j
      \\
      d & c & i
    \end{Bmatrix}
    \,
    \left|
      j_{a c; b d}
    \right\rangle
    \\[2mm]
    \rho(\sigma_3) \,
    \left|i_{a b; c d} \right\rangle
    =
    \left( \lambda_i^{c d} \right)^{-1} \,
    \left|i_{a b; d c} \right\rangle
  \end{gathered}
\end{equation}
See that the quantum $6j$-symbol with the square root term in
$\rho(\sigma_2)$ coincides with the unitary matrix in
Ref.~\citenum{KauffLomon07a}.
Twist operators are 
\begin{equation}
  \rho(\theta_1) \,
  \left| i_{a b; c d} \right\rangle
  =
  \lambda_0^{a a} \,
  \left| i_{a b; c d} \right\rangle
\end{equation}
and so on.

We have
\begin{gather}
  \rho \left(
    \sigma_1 \, \sigma_2 \, \sigma_3^{~2} \, \sigma_2 \, \sigma_1
  \right) \,
  \left| i_{a b; c d} \right\rangle
  =
  \left(
    \lambda_0^{a a}
  \right)^{-2} \,
  \left| i_{a b; c d} \right\rangle
  \\[2mm]
  \rho \left(
    \left(
      \sigma_1 \, \sigma_2 \, \sigma_3
    \right)^4
  \right) \,
  \left| i_{a b; c d} \right\rangle
  =
  \left(
    \lambda_0^{a b} \,
    \lambda_0^{c d}
  \right)^{-2} \,
  \left| i_{a b; c d} \right\rangle
\end{gather}

%%%%%%
\subsection{Braid Relations for Many-Quasi-Particles with Spin-$c/2$}
We study ($n+3$)-quasi-particle with spin-$c/2$.
We set
\begin{equation}
  \label{multi_qubit_c}
  \left| s_1 \, s_2 \, \cdots \, s_n ;c \right\rangle
  =
  \frac{1}{
    \sqrt{N_{s_1 \cdots s_n}^{(c)}}
  }
  \,
  \mbox{
    \raisebox{-.6cm}{
      \includegraphics[scale=0.8]{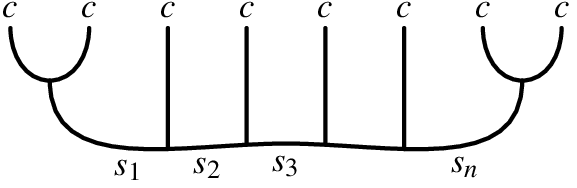}
    }
  }
\end{equation}
where $s_i \geq 0$.
The admissible condition~\eqref{admissible} is read as
\begin{equation}
  \label{admissible_multi-c}
  \begin{gathered}
    c  \leq s_i + s_{i+1} \leq 2 \, K - c
    \\[2mm]
    s_i -c \leq s_{i+1} \leq s_i + c
    \\[2mm]
    s_i + s_{i+1} + c = 0 \mod 2
  \end{gathered}
\end{equation}
where  we mean $s_0 = s_{n+1}=c$.
Thanks to~\eqref{trivalent_theta} the normalization factor is computed as
\begin{equation}
  N_{s_1 \cdots s_n}^{(c)}
  =
  \frac{
    \prod_{i=1}^{n+1} 
    \theta( s_{i-1},  s_i, c)
  }{
    \prod_{i=1}^n
    \Delta_{s_i}
  }
\end{equation}
and we have
\begin{equation}
  \left\langle s_1^\prime  s_2^\prime \cdots s_n^\prime ;c
    \middle|
    s_1 s_2 \cdots s_n ; c
  \right\rangle
  =
  \prod_{i=1}^n \delta_{s_i, s_i^\prime}
\end{equation}
Here the dual  state $
\left\langle s_1 s_2 \cdots s_n ; c\right|$
is defined upside down with same normalization factor.

Actions of the braid operators 
$\sigma_i$
can be computed in essentially same
method with the previous sections.
Results are as follows;
\begin{gather}
  \label{multi_sigma_1}
  \rho(\sigma_1) \, 
  \left| s_1 \cdots s_n ;c \right\rangle
  =
  \left( \lambda_{s_1}^{c c} \right)^{-1} \,
  \left| s_1 \cdots s_n ; c\right\rangle
  \\[2mm]
  \rho(\sigma_{n+2} ) \,
  \left| s_1 \cdots s_n ; c \right\rangle
  =
  \left( \lambda_{s_n}^{c c} \right)^{-1} \,
  \left| s_1 \cdots s_n ; c \right\rangle
\end{gather}
and, for $1<i<n+1$, we have
\begin{multline}
  \label{multi_sigma_3}
  \rho(\sigma_{i+1}) \,
  \left| s_1 \cdots s_n ; c \right\rangle
  \\
  =
  \sum_m
  \sqrt{
    \frac{
      \Delta_{s_i} \, \theta(m, s_{i-1}, c) \, \theta(s_{i+1},m,c)
    }{
      \Delta_{m} \, \theta(s_i, s_{i-1}, c) \, \theta(s_{i+1},s_i,c)
    }
  } \,
  \left(
    \lambda_{s_{i-1}}^{c m}
  \right)^{-1} \,
  \lambda_{s_i}^{s_{i+1} c} \,
  \begin{Bmatrix}
    s_{i-1} & c & m \\
    s_{i+1} & c & s_i
  \end{Bmatrix}
  \\
  \times
  \left| s_1 \cdots m \cdots s_n ; c \right\rangle
\end{multline}
We see that
\begin{gather}
  \rho
  \left(
    \sigma_1 \, \sigma_2 \cdots \sigma_{n+1} \,
    \sigma_{n+2}^{~2} \,
    \sigma_{n+1} \cdots \sigma_2 \,  \sigma_1
  \right)
  \,
  \left| s_1 \cdots s_n ; c \right\rangle
  =
  \left( \lambda_0^{cc} \right)^{-2} \,
  \left| s_1 \cdots s_n ; c \right\rangle  
  \\[2mm]
  \rho\left(
    \left(\sigma_1 \, \sigma_2 \, \cdots \sigma_{n+2} \right)^{n+3}
  \right)
  \,
  \left| s_1 \cdots s_n ; c \right\rangle  
  =
  \left(
    \lambda_0^{c 0}
  \right)^{-2 (n+3)} \,
  \left| s_1 \cdots s_n ; c \right\rangle  
\end{gather}

\subsubsection{Spin-$1/2$ Quasi-Particle States}
In a case of $c=1$,
the admissible condition~\eqref{admissible_multi-c} is
\begin{equation}
  \begin{gathered}
%     s_i = (-1)^{i+1} \mod 2
%     \\[2mm]
    1 \leq s_i + s_{i+1} \leq 2 \, K-1
    \\[2mm]
    s_{i+1} = s_i - 1 
    \qquad
    \text{or}
    \qquad
    s_{i+1} = s_i + 1
%    s_i - 1 \leq s_{i+1} \leq s_i +1
  \end{gathered}
\end{equation}
where  we mean $s_0 = s_{n+1}=1$.
Normalization factor is written with the $\theta$-net,
\begin{equation*}
  \theta(a,b,1)=\Delta_{\max(a,b)}
\end{equation*}
Note that the notation
$\left| s_1 s_2 s_3 ;1 \right\rangle$ is different
from~\eqref{base_6-particle} used in 6-quasi-particle case;
$\left| 010 ;1\right\rangle$,
$\left| 012 ;1\right\rangle$, and
$\left| 210 ;1\right\rangle$ respectively coincide with
$\left| 00 \right\rangle$,
$\left| 01 \right\rangle$, and
$\left| 10 \right\rangle$ in~\eqref{base_6-particle}.
Remaining two states,
$\left| 212 ;1\right\rangle$
and
$\left| 232 ;1\right\rangle$,
are related to
$\left| 11 \right\rangle$
and
$\left| C \right\rangle$ in~\eqref{base_6-particle}
through the $F$-matrix~\eqref{F_1212}.

The braid matrices follows
from~\eqref{multi_sigma_1}--\eqref{multi_sigma_3}, and
non-zero quantum $6j$-symbols therein are
\begin{equation*}
  \begin{aligned}
    \begin{Bmatrix}
      a & 1 & a+1
      \\
      a+2 & 1 & a+1
    \end{Bmatrix}
    & =
    \begin{Bmatrix}
      a & 1 & a+1
      \\
      a & 1 & a-1
    \end{Bmatrix}
    =
    \begin{Bmatrix}
      a & 1 & a-1 
      \\
      a-2 & 1 & a-1
    \end{Bmatrix}
    =
    1
    \\[2mm]
    \begin{Bmatrix}
      a & 1 & a-1
      \\
      a & 1 & a +1
    \end{Bmatrix}
    & =
    \frac{
      \Delta_{a-1} \, \Delta_{a+1}
    }{
      \left( \Delta_a \right)^2
    }
    \\[2mm]
    \begin{Bmatrix}
      a & 1 & a \pm 1
      \\
      a & 1 &  a \pm 1
    \end{Bmatrix}
    & =
    \mp \frac{1}{\Delta_a}
  \end{aligned}
\end{equation*}

%%%%
\subsubsection{Spin-$1$ Quasi-Particle States}
The admissible condition~\eqref{admissible_multi-c} with $c=2$ is read as
\begin{equation}
  \begin{gathered}
    2 \leq s_i + s_{i+1} \leq 2 \,K-2
    \\[2mm]
    s_{i+1} \in \{ s_i \pm 2, s_i \}
  \end{gathered}
  \label{admissible_many_spin-1}
\end{equation}
and $s_0=s_{n+1}=2$.
The normalization factor is calculated  by use of the following
$\theta$-net;
\begin{gather*}
  \theta(2 \, a, 2 \, b, 2)
  =
  \begin{cases}
    \frac{
      \Delta_{2a} \, \Delta_{2a+1}
    }{
      d \, \Delta_{2a-1}
    }
    & \text{for $a=b$}
    \\[2mm]
    \Delta_{2 \max(a,b)}
    & \text{for $|a-b|=1$}
    \\[2mm]
    0
    & \text{others}
  \end{cases}
%   \theta(2 \, a , 2 \, a , 2)
%   =
%   \frac{
%     \Delta_{2a} \, \Delta_{2a+1}
%   }{
%     d \, \Delta_{2a-1}
%   }
%   \\[2mm]
% %
%   \theta(2 \, a , 2 \, a - 2 , 2)
%   =
%   \Delta_{2a}
\end{gather*}
States,
$\left|
  \frac{s_1}{2}
\right\rangle$~\eqref{1-qubit_spin_1},
and
$\left|
  \frac{s_1}{2}
  \frac{s_2}{2}
\right\rangle$~\eqref{state_5_quasi},
respectively correspond to
$\left|
  s_1 
  ;2
\right\rangle$,
and
$\left|
  s_1 s_2;
  2
\right\rangle$.
Non-zero $6j$-symbols are;
\begin{equation*}
  \begin{aligned}
    \begin{Bmatrix}
      2 \,a & 2  & 2 \,a \pm 2
      \\
      2 \, a \pm 4 & 2 & 2 \, a \pm 2
    \end{Bmatrix}
    & =
    \begin{Bmatrix}
      2 \,a & 2  & 2 \,a + 2
      \\
      2 \, a +2 & 2 & 2 \, a
    \end{Bmatrix}
    =
    1
    \hspace{-28mm}
    \\[2mm]
    \begin{Bmatrix}
      2 \,a & 2  & 2 \,a + 2
      \\
      2 \, a +2 & 2 & 2 \, a+2
    \end{Bmatrix}
    & =
    \frac{d}{\Delta_{2a+1}}
    &
    \begin{Bmatrix}
      2 \,a & 2  & 2 \,a 
      \\
      2 \, a +2 & 2 & 2 \, a+2
    \end{Bmatrix}
    & =
    \frac{
      \Delta_{2a-1} \, \Delta_{2a+3}
    }{
      \left( \Delta_{2a+1} \right)^{2}
    }
    \\[2mm]
    \begin{Bmatrix}
      2 \,a & 2  & 2 \,a 
      \\
      2 \, a \pm 2 & 2 & 2 \, a
    \end{Bmatrix}
    & =
    \frac{d}{\Delta_{2a \pm 1}}
    &
    \begin{Bmatrix}
      2 \,a & 2  & 2 \,a \pm 2
      \\
      2 \, a  & 2 & 2 \, a \pm 2
    \end{Bmatrix}
    & =
    \frac{d}{\Delta_{2a \pm 1} \, \Delta_{2a}}
    \\[2mm]
    \begin{Bmatrix}
      2 \,a & 2  & 2 \,a  -2
      \\
      2 \, a  & 2 & 2 \, a
    \end{Bmatrix}
    & =
    \frac{\Delta_{2a-2} \, \Delta_{2a+1}}{
      \left( \Delta_{2a-1} \right)^2 \, \Delta_{2a}
    }
    &
    \begin{Bmatrix}
      2 \,a & 2  & 2 \,a  -2
      \\
      2 \, a  & 2 & 2 \, a+2
    \end{Bmatrix}
    & =
    \frac{\Delta_{2a-2} \, \Delta_{2a+2}}{
      \left( \Delta_{2a} \right)^2
    }
    \\[2mm]
    \begin{Bmatrix}
      2 \,a & 2  & 2 \,a  
      \\
      2 \, a  & 2 & 2 \, a-2
    \end{Bmatrix}
    & =
    \frac{d^2}{\Delta_{2a+1}}
    &
    \begin{Bmatrix}
      2 \,a & 2  & 2 \,a  
      \\
      2 \, a  & 2 & 2 \, a
    \end{Bmatrix}
    & =
    \frac{
      \Delta_{2a-2} \, \Delta_{2a+2} -1}{
      \Delta_{2a-1} \, \Delta_{2a+1}
    }    
    \\[2mm]
    \begin{Bmatrix}
      2 \,a & 2  & 2 \,a 
      \\
      2 \, a  & 2 & 2 \, a+2
    \end{Bmatrix}
    & =
    -
    \frac{d^2 \, \Delta_{2a-1} \, \Delta_{2a+2}}{
      \left( \Delta_{2a+1} \right)^2 \, \Delta_{2a}
    }
    &
    \begin{Bmatrix}
      2 \,a & 2  & 2 \,a  +2
      \\
      2 \, a  & 2 & 2 \, a
    \end{Bmatrix}
    & =
    - \frac{1}{\Delta_{2a-1}}
    \\[2mm]
    \begin{Bmatrix}
      2 \,a & 2  & 2 \,a  +2
      \\
      2 \, a  & 2 & 2 \, a-2
    \end{Bmatrix}
    & =
    \begin{Bmatrix}
      2 \,a & 2  & 2 \,a  
      \\
      2 \, a -2  & 2 & 2 \, a- 2
    \end{Bmatrix}
    = 1
    \hspace{-28mm}
    \\[2mm]
%
%     \begin{Bmatrix}
%       2 \,a & 2  & 2 \,a  +2
%       \\
%       2 \, a  & 2 & 2 \, a+2
%     \end{Bmatrix}
%     & =
%     \frac{d}{\Delta_{2a} \, \Delta_{2a+1}}
%     &
%
    \begin{Bmatrix}
      2 \,a & 2  & 2 \,a  -2
      \\
      2 \, a  -2 & 2 & 2 \, a-2
    \end{Bmatrix}
    & =
    - \frac{1}{\Delta_{2a-1}}
%    \\[2mm]
%
%     \begin{Bmatrix}
%       2 \,a & 2  & 2 \,a 
%       \\
%       2 \, a  -2 & 2 & 2 \, a
%     \end{Bmatrix}
%     & =
%     \frac{d}{\Delta_{2a-1}}
     &
    \begin{Bmatrix}
      2 \,a & 2  & 2 \,a  -2
      \\
      2 \, a -2 & 2 & 2 \, a
    \end{Bmatrix}
    & =
    \frac{
      \Delta_{2a-3} \, \Delta_{2a+1}
    }{
      \left( \Delta_{2a-1} \right)^2
    }    
  \end{aligned}
\end{equation*}

It should be remarked that,
in Ref.~\citenum{FTLTKWF07a}, studied was the quantum spin Hamiltonian
acting on state~\eqref{multi_qubit_c} with spin-$1$ for the $SU(2)_3$ theory.

\subsection{Topological Entanglement Entropy of Many-Quasi-Particle States}
\label{sec:topological_entropy}

We first discuss the entanglement entropy of
the 4-quasi-particle states
$\left| i_{ab;cd} \right\rangle$~\eqref{4_qp_general}.
We set   a state by
\begin{equation}
  \label{general_Psi_4}
  \left| \Psi \right\rangle
  = \sum_i p_i \, \left| i_{ab;cd} \right\rangle
\end{equation}
where $p_i \in \mathbb{C}$ and
\begin{equation*}
  \sum_i \left| p_i \right|^2 =1
\end{equation*}
Note that the sum of $i$ runs over such that triples $(i,a,b)$ and
$(i,c,d)$ are admissible~\eqref{admissible}.
The density matrix is written as
\begin{equation}
  \rho =
  \left| \Psi \right\rangle \,
  \left\langle \Psi \right|
  =
  \sum_{i,j}
  p_i \, p_j^* \,
  \sqrt{
    \frac{\Delta_i}{
      \theta(a,b,i) \, \theta(c,d,i)} \,
    \frac{\Delta_j}{
      \theta(a,b,j) \, \theta(c,d,j)} \,
  }  \,
  \mbox{
    \raisebox{-9mm}{
      \includegraphics[scale=0.8]{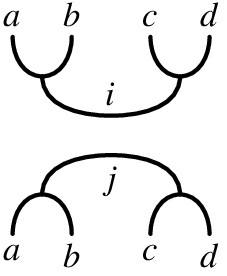}
    }
  }
\end{equation}

We first consider the entanglement between $A$ and $B$ when
Alice owns  spin-$a/2$ and -$b/2$ quasi-particles and Bob others;
we write
$A=\{a,b\}$ and
$B=\{c,d\}$.
Alice's reduced density matrix  is depicted as
\begin{align}
  \rho_{\{a , b \}}
  &=
  \sum_{i,j}
  p_i \, p_j^* \,
  \sqrt{
    \frac{\Delta_i}{
      \theta(a,b,i) \, \theta(c,d,i)} \,
    \frac{\Delta_j}{
      \theta(a,b,j) \, \theta(c,d,j)} \,
  }  \,
  \mbox{
    \raisebox{-11mm}{
      \includegraphics[scale=0.8]{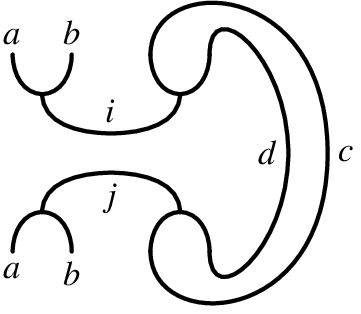}
    }
  }
  \nonumber \\
  & =
  \sum_i
  \left| p_i \right|^2 \,
  \frac{1}{\theta(a,b,i)} \,
  \mbox{
    \raisebox{-8mm}{
      \includegraphics[scale=0.8]{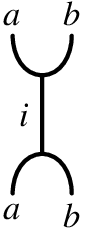}
    }
  }
  \label{reduced_density_a_b}
\end{align}
where we have used~\eqref{trivalent_theta} in the second equality.
Then the entanglement entropy~\eqref{von_Neumann_entropy} is computed
from~\eqref{depict_S_A} as
\begin{equation}
  S_{\{a , b \}}
  =
  - \sum_i
  \left| p_i \right|^2
  \log \left(
    \frac{
      \left| p_i \right|^2}{
      \Delta_i}
  \right)
\end{equation}
where we have used~\eqref{trivalent_theta} again.
The identity~\eqref{trivalent_theta} indicates
that 
the expression~\eqref{4_qp_general} has orthonormal bases in 
Alice's and Bob's spaces.
Then we obtain the topological entanglement
entropy~\eqref{topological_entanglement} as
\begin{equation}
  S_{\{a , b \}}^{\text{topo}}
  =
  \sum_i
  \left| p_i \right|^2
  \log \Delta_i
\end{equation}
Indeed $\Delta_i$ denotes the quantum dimension $d_i$ of the quasi-particle
with spin-$i/2$~\eqref{quantum_dimension}
who  intertwines Alice
$\{a,b\}$ and Bob $\{c,d\}$.

To study a case when Alice has spin-$b/2$ and -$c/2$ quasi-particles,
$A=\{b,c\}$ and
$B=\{a,d \}$,
we prepare  different bases using the quantum $6j$-symbol~\eqref{6j_symbol}.
We have
\begin{equation}
  \label{general_Psi_4_2}
  \left| \Psi \right\rangle
  =
  \sum_k
  \widetilde{p_k} \,
  \left| \widetilde{k_{ab:cd}} \right\rangle
\end{equation}
where
\begin{gather}
  \label{general_different_base}
  \left| \widetilde{k_{ab:cd}} \right\rangle
  =
  \sqrt{
    \frac{\Delta_k}{\theta(k,b,c) \, \theta(k,a,d)}
  } \,
  \mbox{
    \raisebox{-0.3cm}{
      \includegraphics[scale=0.8]{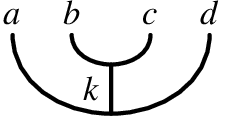}
    }
  }
  \\[2mm]
  \widetilde{p_k}
  =
  \sum_i p_i \,
  \sqrt{
    \frac{\Delta_i \, \theta(b,c,k) \, \theta(a,d,k)}{
      \Delta_k \, \theta(a,b,i) \, \theta(c,d,i)}
  } \,
  \begin{Bmatrix}
    a & b & k \\
    c & d & i
  \end{Bmatrix}
\end{gather}
Using
$
\begin{Bmatrix}
  a & b & i \\
  c & d & j
\end{Bmatrix}^*
=
\begin{Bmatrix}
  b & a & i \\
  d & c & j
\end{Bmatrix}$,
the orthogonal identities~\eqref{orthogonal_6j},  and
symmetries~\eqref{symmetry_6j} and~\eqref{symmetry_6j_2},
we indeed see that
\begin{equation*}
  \sum_k \left| \widetilde{p_k} \right|^2 = 1
\end{equation*}
Then
one sees that
Alice's reduced density matrix is computed as
\begin{align}
  \rho_{\{ b, c\}}
  & =
  \sum_{j,k}
  \widetilde{p_k} \, \widetilde{p_j}^* \,
  \sqrt{
    \frac{\Delta_k}{
      \theta(b,c,k) \, \theta(a,d,k)} \,
    \frac{\Delta_j}{
      \theta(b,c,j) \, \theta(a,d,j)} \,
  }  \,
  \mbox{
    \raisebox{-.9cm}{
      \includegraphics[scale=0.8]{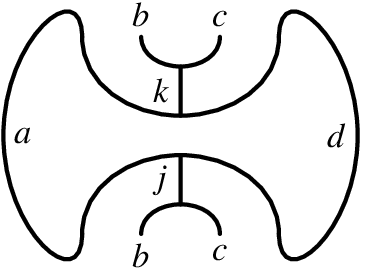}
    }
  }
  \nonumber
  \\
  & =
  \sum_k
  \left| \widetilde{p_k} \right|^2 \,
  \frac{1}{\theta(b,c,k)} \,
  \mbox{
    \raisebox{-8mm}{
      \includegraphics[scale=0.8]{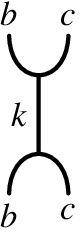}
    }
  }
\end{align}
which gives
\begin{equation}
  S_{\{b , c \}}
  =
  - \sum_k
  \left| \widetilde{p_k} \right|^2
  \log \left(
    \frac{
      \left| \widetilde{p_k} \right|^2}{
      \Delta_k}
  \right)
\end{equation}
By the same argument, we can identify
the topological entanglement entropy with
\begin{equation}
    S_{\{b, c \}}^{\text{topo}}
    =
    \sum_k
    \left| \widetilde{p_k} \right|^2
    \log \Delta_k
\end{equation}
which shows that we have the quantum dimension of quasi-particle which
intertwines Alice and Bob in the
expression~\eqref{general_different_base}.

It is straightforward to study a case when Alice has only a spin-$a/2$
quasi-particle;
$A=\{a\}$
and
$B=\{b,c,d\}$.
From~\eqref{reduced_density_a_b},
we have Alice's reduced density matrix as
\begin{equation}
  \rho_{\{ a \}}
  =
  \frac{1}{\Delta_a} \,
  \mbox{
    \raisebox{-.7cm}{
      \includegraphics[scale=0.8]{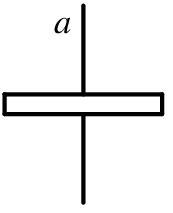}
    }
  }
\end{equation}
which gives
\begin{equation}
  S_{ \{ a \} }^{\text{topo}}
  =
  S_{ \{ a \} }
  =
  \log \Delta_a
\end{equation}

As the entanglement entropy~\eqref{von_Neumann_entropy}
satisfies~\eqref{entropy_A_B}, we may expect
\begin{equation}
  \label{S_abc_topology}
  S_{\{a, b, c\}}^{\text{topo}}
  =
  S_{\{a, b, c\}}
  = \log \Delta_d
\end{equation}
We shall check this formula skein-theoretically.
The reduced density matrix is given by
\begin{equation}
  \rho_{\{a, b, c \}}
  =
  \sum_{i,j} 
  m_{i,j} \,
  \mbox{
    \raisebox{-10mm}{
      \includegraphics[scale=0.8]{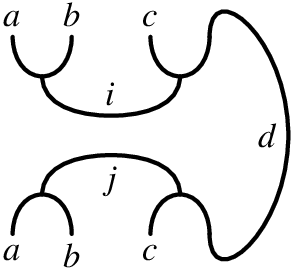}
    }
  }
\end{equation}
where
\begin{equation}
  m_{i,j} =
  p_i \, p_j^* \,
  \sqrt{
    \frac{\Delta_i}{
      \theta(a,b,i) \, \theta(c,d,i)} \,
    \frac{\Delta_j}{
      \theta(a,b,j) \, \theta(c,d,j)} \,
  }  
\end{equation}
Then we have
\begin{align*}
  \Tr_{\{a, b, c \}}
  \left(
    \rho_{\{a, b, c \}}
  \right)^{n}
  & =
  \sum_{j_1,\dots, j_n}
  m_{j_n,j_1} \, m_{j_1, j_2} \cdots m_{j_{n-1},j_n} \,
  \left(
    \prod_{k=1}^n
    \frac{\theta(j_k,a,b) \, \theta(j_k,c,d)}{
      \Delta_{j_k} \, \Delta_d
    }
  \right) \,
  \Delta_d
  \\
  & =
  \sum_{j_1,\dots, j_n}
  \left| p_{j_1} \right|^2 \cdots
  \left| p_{j_n} \right|^2 \,
  \frac{1}{
    \Delta_d^{~n-1}
  }
  \\
  &=
  \frac{1}{\Delta_d^{~n-1}}
\end{align*}
The replica trick~\eqref{entropy_replica} gives~\eqref{S_abc_topology}
as we expected.

We stress again the the topological entanglement entropy 
$S_A^{\text{topo}}$
also depends
on how to define $A$ and $B$.

To conclude we have shown that the topological entanglement entropy is
written in terms of the quantum dimension.
Namely in the case that
the  state
$\left|\Psi\right\rangle$ with non-Abelian quasi-particles
has a form~\eqref{pure_Psi},
the entanglement entropy is modified to
\begin{equation}
  S_A=
  \sum_j 
  \left| p_j \right|^2 \log
  \left( \frac{d_j}{
      \left| p_j \right|^2 
    }
  \right)
\end{equation}
which includes 
the topological entanglement entropy
\begin{equation}
  \label{formula_topology}
  S_A^{\text{topo}}
  =
  \sum_j
  \left| p_j \right|^2
  \log d_j
\end{equation}
where $\left| p_j \right|^2$ denotes a probability,
$\sum_j \left| p_j \right|^2 =1$,
that, in the state $\left| \Psi \right\rangle$, quasi-particle with
quantum
dimension $d_j$ 
plays a role of Cupid who intertwines Alice and Bob.
% We mean that
% $d_j$ denotes the quantum dimension of quasi-particle which
% intertwines  Alice and Bob.
We have obtained~\eqref{formula_topology} for 4-quasi-particle states,
but a computation for many-quasi-particle states is same
because
recursive use of the identity~\eqref{null_6j} enables us to rewrite
any braided trivalent graphs into a  graph where only a single Wilson
line intertwines  Alice and Bob.
Once we have such trivalent graph,
the identity~\eqref{trivalent_theta} proves that we indeed have
orthogonal bases in a sense of~\eqref{orthogonal_base}.
So our 
formula~\eqref{formula_topology} for the topological entanglement
entropy is correct for other many-quasi-particle states.

%%%%%
\section{Concluding Remarks and Discussions}

%%%%
%\section{Quantum Compiling}

We have studied the topological properties
of many-quasi-particle states in the Read--Rezayi state whose
effective theory is the $SU(2)_K$ Chern--Simons theory.
Applying  the
skein theory, we have obtained the unitary representation of the
braiding matrices of quasi-particle states.
These braiding matrices can be used as 
fundamental gates to construct other unitary gates.
For example, by use of
braiding matrices~\eqref{braiding_4-qp_Pfaffian}
of the 4-quasi-particle states in the Pfaffian state,
we can construct
the NOT gate and the Hadamard gate by~\cite{LGeorg06b}
\begin{gather*}
  \rho\left(
    \theta_1^{~2} \,
    \sigma_2^{~2}
  \right)
  =
  \begin{pmatrix}
    0 & 1 \\
    1 & 0
  \end{pmatrix}
  \\[2mm]
  \rho\left(
    \theta_1 \,
    \sigma_1 \, \sigma_2 \, \sigma_1
  \right)
  =
  \frac{-1}{\sqrt{2}} \,
  \begin{pmatrix}
    1 & 1 \\
    1 & -1
  \end{pmatrix}
\end{gather*}
Unfortunately as was shown in
Refs.~\citenum{FreeLarsWang02a,FreeLarsWang02b} this $SU(2)_2$ theory
is not universally computable, although
proposed in Ref.~\citenum{SBravy06a}
to use another non-topological gate for universal computations.
The simplest non-Abelian theory which is responsible for universal
computation is the Fibonacci anyon model, \emph{i.e.}, $SU(2)_3$
theory.
The Solovay--Kitaev 
algorithm~\cite{NielsChuan00Book,KitaShenVyal02a}
supports that any unitary gates can be constructed from braiding
matrices~\eqref{braiding_Fibonacci_1}
and~\eqref{braiding_Fibonacci_2}
of the Fibonacci anyons, although
it was shown~\cite{FreeWang06a} that the gate of  large Fourier
transformation cannot be realized exactly. 
In fact
several unitary gates were approximated
using these braiding
matrices~\cite{HorZikBonSim07a,SiBoFrPeHo06a,BoneHormZikoSimo05a}.
It may be interesting to construct fundamental gates such as NOT,
Hadamard, and CNOT,
using the braiding matrices of $SU(2)_K$ theory presented
here.

Using the  braided trivalent graph as bases of Hilbert space of
many-quasi-particle states,
we have proposed a method to compute the bipartite entanglement entropy.
Topological effect to the entanglement entropy was discussed in
Refs.~\citenum{KitaePresk06a,LeviXWen06a},  but extraction of
topological part from the entanglement entropy is generally subtle in
computations in physical models such as the quantum dimer
model~\cite{FurukMisgu07a,PapaRamaFrad07a}.
Our construction   is suitable  to avoid such subtlety due to that the
skein theory
or the Chern--Simons theory is purely topological gauge theory.
We have indeed  obtained exactly
topological contribution to the entanglement entropy.
We have shown that many-quasi-particle states have the
topological entanglement entropy~\eqref{formula_topology}
which depends only on the quantum dimension of a 
quasi-particle intertwining Alice and Bob.

\section*{Acknowledgments}
This work is supported in part  by Grant-in-Aid 
from the Ministry of Education, Culture, Sports, Science and
Technology of Japan.

%%%%%%%%%%%%%%

%%%%%%%%%%%%%% %ewpage

\end{document}